\documentclass[12pt,preprint]{aastex}

\shorttitle{Quasar Microlensing with Extended Sources} 
\shortauthors{Mortonson et al.}

\begin{document}

\title{Size is Everything: Universal Features of Quasar Microlensing
with Extended Sources}

\author{Michael J. Mortonson\altaffilmark{1,2} and Paul L. Schechter}
\affil{Department of Physics, Massachusetts Institute of Technology, 77 Massachusetts Avenue, Cambridge, MA 02139}

\and

\author{Joachim Wambsganss\altaffilmark{3}}
\affil{Universit\"{a}t Potsdam, Institut f\"{u}r Physik, Am Neuen Palais 10, 14467 Potsdam, Germany}

\altaffiltext{1}{\emph{mjmort@uchicago.edu}}
\altaffiltext{2}{present address: Department of Physics, University of Chicago, 
5640 S. Ellis Avenue, Chicago, IL 60637}
\altaffiltext{3}{present address: Astronomisches Rechen-Institut, Universit\"{a}t 
Heidelberg, Moenchhofstr. 12-14, 69120 Heidelberg, Germany}

\begin{abstract}
We examine the effect that the shape of the source brightness profile has on
the magnitude fluctuations of images in quasar lens systems due to microlensing.
 We do this by convolving a variety of accretion disk models (including
Gaussian disks, uniform disks, ``cones," and a Shakura-Sunyaev
thermal model) with two
magnification maps in the source plane, one with convergence $\kappa=0.4$
 and shear $\gamma=0.4$ (positive parity), and the other with $\kappa=\gamma=0.6$
(negative parity).  By looking
at magnification histograms of the convolutions and using chi-squared tests to
determine the number of observations that would be necessary to distinguish
histograms associated with different disk models, we find that, for circular
disk models, the
microlensing fluctuations are relatively insensitive to all properties
of the models except the half-light radius of the disk. 
Shakura-Sunyaev models are sufficiently well constrained by observed 
quasar properties that we can estimate the half-light radius at optical 
wavelengths for a typical quasar.   If Shakura-Sunyaev models are 
appropriate, the half-light radii are very much smaller than 
the Einstein rings of intervening stars and the quasar can be reasonably 
taken to be a point source except in the immediate vicinity of 
caustic crossing events.
\end{abstract}


\section{INTRODUCTION} \label{fact}

The deflection angles associated with gravitational microlensing 
of quasars, due to 
stellar-mass objects such as stars in a lensing galaxy, are on the order 
of 1 microarcsecond, too small to be resolved into separate microimages.  
However, microlensing can have significant effects on the magnitudes of 
macroimages.  Magnitude fluctuations from microlensing have been detected 
in several lensed quasars.  These effects were first observed by~\citet{i89} 
in the quasar Q2237+0305.  This quasar has been recently monitored as 
part of the Optical Gravitational Lensing Experiment (OGLE), and 
microlensing fluctuations with amplitudes up to 1.3 magnitudes in a two-year 
period have 
been observed~\citep{w00a,w00b}.  Microlensing can be distinguished from 
intrinsic quasar time-variability by looking for fluctuations that are 
uncorrelated between the macroimage light curves.  Quasar microlensing 
could help explain observed flux ratio anomalies for quasars in which the 
magnitude differences between the macroimages differ greatly from those 
predicted by theory~\citep{wms95,mm01,c02,dk02,sw02}.

There is a large number of parameters that could be important for modeling 
lens systems: properties of the source, including its size and shape; lens 
properties such as the mass distribution of objects that make up the lens; and 
cosmological parameters like the Hubble constant.  
  Although we expect all of these properties to affect the physics of lensing 
in some way, the effects of some properties are more significant than the 
effects of others.
  It is important to find out which
parameters have little effect on the observables in lensing so that
those properties can be neglected in lens models.

For quasar microlensing, there is a great deal of evidence that the size of 
the source has a large effect on the fluctuations due to
microlensing when the quasar crosses a caustic in the source plane.  
Observations of extragalactic microlensing have
 been used to place constraints on the sizes
of quasars and on the scales over which different quasar emission mechanisms 
operate~\citep[e.g.,][]{wwt00,y01,sg02,w02,su03}.
 A large extended source covers more microlensing caustics in the source plane at any given 
time than a small source, so its brightness varies less as it moves relative to the lens and
observer.  As a general rule, the variability of a lensed source will only be significantly affected
by microlensing if the source is smaller than the projection of the Einstein radius
of a microlens into the source plane~\citep{cm02}.  (Note, however, 
that \cite{rs97} have argued that in some circumstances, 
even relatively large sources can have significant 
fluctuations due to microlensing.)  

The same effect could be responsible for differences between emission-line and continuum
flux ratios, which have been found
in a number of lens systems~\citep[e.g.,][]{w93,s98,b02,w03,m03,c04}.
 A possible explanation for these differences is that the broad emission
line regions of quasars are much larger than the Einstein radii of the microlenses,
and
 the continuum-emitting regions are much smaller than the Einstein radii~\citep{mm03}.

The dependence of temperature on radius in quasar accretion disks also leads
to size-dependent effects.  Since
the disk is cooler far from the center than it is near the central black
hole, the disk will have a larger effective
radius when observed at long wavelengths than it will when observed at
short wavelengths~\citep{v03}.  At long wavelengths,
therefore, we expect the magnitude variations due to microlensing to be suppressed.
The Shakura-Sunyaev accretion disk model that we use (Section~\ref{adss})
incorporates the temperature profile of
the disk so that we can study the effects 
of varying wavelength and source size on microlensing fluctuations.
  Besides using photometric observations of microlensing, it has also been suggested
that astrometric observations, looking for small shifts in image positions due
to microlensing, could constrain the sizes of quasars at different
wavelengths~\citep{li98,tw04}.

If we describe the size of a source by its half-light radius ($r_{1/2}$), the radius at
which half of the light is interior to the radius and half of it is outside, then
we can construct different source models with the same half-light radii but
with their brightness distributed in the source plane in different ways.
 We will refer to this
distribution of brightness as the ``shape" of the brightness profile.
 Note that all of the source models we consider here are
circularly symmetric, so ``shape" does not refer to the shape of the contours
of constant brightness in the source plane, but rather how the spacing of those contours varies
with radius (i.e., the one-dimensional surface brightness profile).

  The
question we would like to address is this: for sources with the same size,
as determined by the half-light radius, to what extent does the shape of
each source influence the fluctuations due to microlensing of the source?
The answer to this question tells us how important the shape of the
source brightness profile is to observations and models of microlensing.

\citet{ak99} and~\citet{w00} have also looked at the connection between source properties
and microlensing, but their studies use a large number of parameters for the disk
models, whereas our models have a small number of parameters while still covering 
a wide range of disk shapes.  \citet{k03} uses disk models similar to our 
Shakura-Sunyaev model.  These studies and others \citep[e.g.,][]{g88,my99} use 
microlensing light curves and caustic-crossing events to infer source properties.  
\citet{dk05} examine finite sources in \emph{milli}lensing by finding their 
effect on image 
positions and magnifications.  In contrast to these studies, 
our main tool for analyzing the relation between source properties and 
microlensing fluctuations is the magnification histogram.

\section{ACCRETION DISK MODELS} \label{diskmod}

To study the effects of the shape of a source brightness profile on
microlensing fluctuations, we use a variety of highly idealized accretion disk
models with different shapes to model the source quasar.  The first three
models (Sections~\ref{adg} to~\ref{adc}) are adopted not because they are 
necessarily realistic, but because
they are mathematically simple and span a wide range of possibilities.  The
fourth model (Section~\ref{adss}), while still an idealization, is physically 
motivated.

\subsection{Gaussian Disks} \label{adg}

One common type of accretion disk model is a circular two-dimensional
Gaussian~\citep[e.g.,][]{w02}.  The surface brightness profile (with units of
erg~s$^{-1}$~cm$^{-2}$~sr$^{-1}$) can be written
\begin{equation}
G(r)=F\frac{D_S^2}{2\pi \sigma^2}e^{-r^2/2\sigma^2}, \label{gauss}
\end{equation}
where $F$ is the total flux at Earth from the disk (with units of erg~s$^{-1}$~cm$^{-2}$), 
$D_S$ is the distance from Earth to the quasar, $r$ is the radius
in the source plane from the center of the disk, and $\sigma$ is the
width of the Gaussian (with units of length, measured in the source plane).

\subsection{Uniform Disks} \label{adu}

Even less realistic than the Gaussian disk, a uniform disk is
 the simplest disk model imaginable.  The uniform disk model has a
surface brightness of $FD_S^2/(\pi R^2)$ for radii $0~<~r~<~R$ (with $F$  
and $D_S$ as defined above), and is zero for $r>R$.

\subsection{Cones} \label{adc}

The ``cone" disk model is peaked at the center, and decreases linearly
with increasing radius until it reaches zero at a radius $R$, outside of
which the model is zero everywhere.  The surface brightness profile is
\begin{equation}
C(r)=F\frac{3D_S^2}{\pi R^2}\left(1-\frac{r}{R}\right),~r<R, \label{cone}
\end{equation}
 where $r$ is the radius from the center, and $F$ and $D_S$ are the same as above.

\subsection{Shakura-Sunyaev Disks} \label{adss}

The last accretion disk model we consider is a thin static
disk, viewed face-on, with a two-dimensional brightness profile
determined by the temperature at each part of the disk, as in several other 
microlensing studies~\citep[e.g.,][]{y98,y99,t01,k03}. Though more
complicated than the previous models, it is still simpler than the similar
thermal disk models used by~\citet{ak99} and~\citet{w00}.
Many of the results we present in Section~\ref{histchap} use this disk model.
In Section~\ref{typq} we relate the properties of this disk model to 
physical quantities for typical quasars.

We begin with a temperature-radius relation for the disk adapted from 
\citet{ss73}:
\begin{equation}
T(r) = 2.049T_0\left(\frac{r_{in}}{r}\right)^{3/4}\left(1-\sqrt{\frac{r_{in}}{r
}}\right)^{1/4}, \label{temp}
\end{equation}
where $T_0$ is the peak
disk temperature, and $r_{in}$ is the radius
of the inner edge of the accretion disk,
which we take to be the radius of the innermost stable circular
orbit around the central black hole.  Thus, $r_{in}$
depends on the black hole mass.

We assume that the disk radiates as a black body with a monochromatic
specific intensity $B_{\lambda}(T)$ that depends
 on the temperature, and therefore on the radius. (All wavelengths are
assumed to be in the quasar frame,
so to compare with wavelengths in the observer's frame the quasar's
redshift must be accounted for.)  Using Equation~(\ref{temp}),
we can write the specific intensity as a function of radius:
\begin{equation}
B_{\lambda}(r) = \frac{2 hc^2}{\lambda^5} \left\{\exp\left[0.488\frac{hc}
{\lambda kT_0} \left(\frac{r}{r_{in}}
\right)^{3/4}\left(1-\sqrt{\frac{r_{in}}{r}}\right)^{-1/4}\right] - 1\right\}^{
-1}. \label{bb1}
\end{equation}

It is convenient to use dimensionless variables for the parameters, so we
define a dimensionless wavelength, $x$, and a dimensionless radius, $s$:
\begin{equation}
x\equiv \frac{kT_0}{hc} \lambda,~s\equiv \frac{r}{r_{in}}, \label{dimless}
\end{equation}
which makes the specific intensity
\begin{equation}
B_x(s) = \frac{a}{x^5} \left\{\exp\left[\frac{0.488}{x}\left(\frac{s^3}{1-s^{-1
/2}}\right)^{1/4}\right]-1\right\}^{-1}, \label{bb2}
\end{equation}
where we define $a\equiv 2r_{in}^2h^{-3}c^{-2}(kT_0)^4$.
For the maximum disk temperature $T_0$ (at $r=1.36r_{in}$), the peak
 of $B_x(s)$ is at $x_0 = 0.2014$.

Since the disk radiates at cooler temperatures with increasing distance from 
the center, observations at different
wavelengths will detect different parts of the disk~\citep{wp91, gm97}.  
To take the wavelength dependence into
account, we define a set of filters associated with
specific ranges of the dimensionless wavelength $x$.  The filter 
numbers increase
 with increasing wavelength, with filter~0
centered at $x=x_0$.  The ranges of $x$ are chosen so that
the filters span the space of wavelengths without overlapping
(that is, $x_{i,max} = x_{i+1,min}$ where $x_{i,min}$
and $x_{i,max}$ are the minimum and maximum wavelengths for filter~$i$).
We assume that each filter transmits $100\%$ of the light in its
wavelength range.  The filters have constant $\Delta(\log x)=\Delta x/x_i=1/5$,
 so
\begin{equation}
x_i \approx e^{0.2i-1.6025}, \label{filter}
\end{equation}
where $x_i$ is the central wavelength of filter $i$.

\begin{figure}
\epsscale{0.5}
\plotone{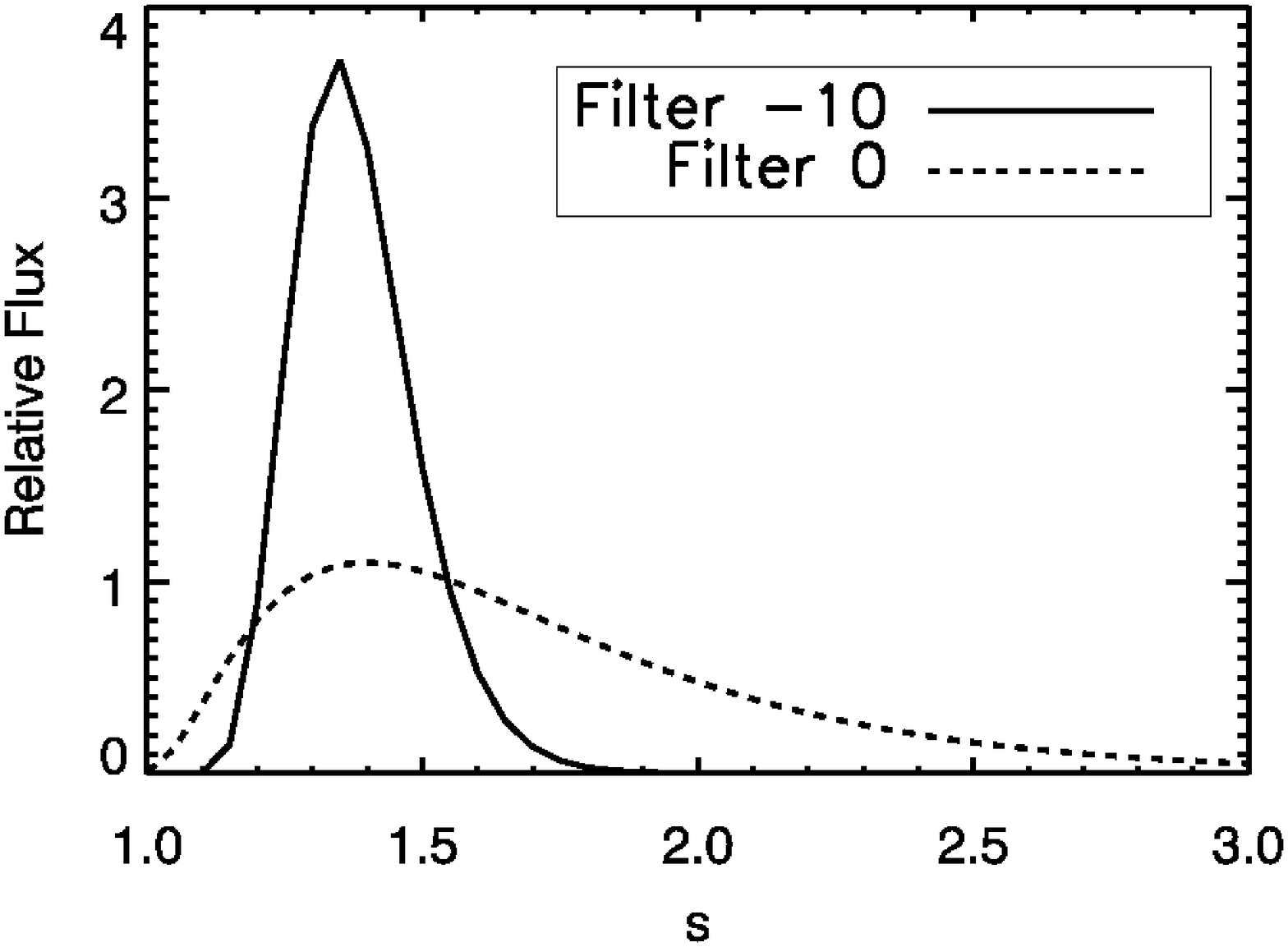}

\plotone{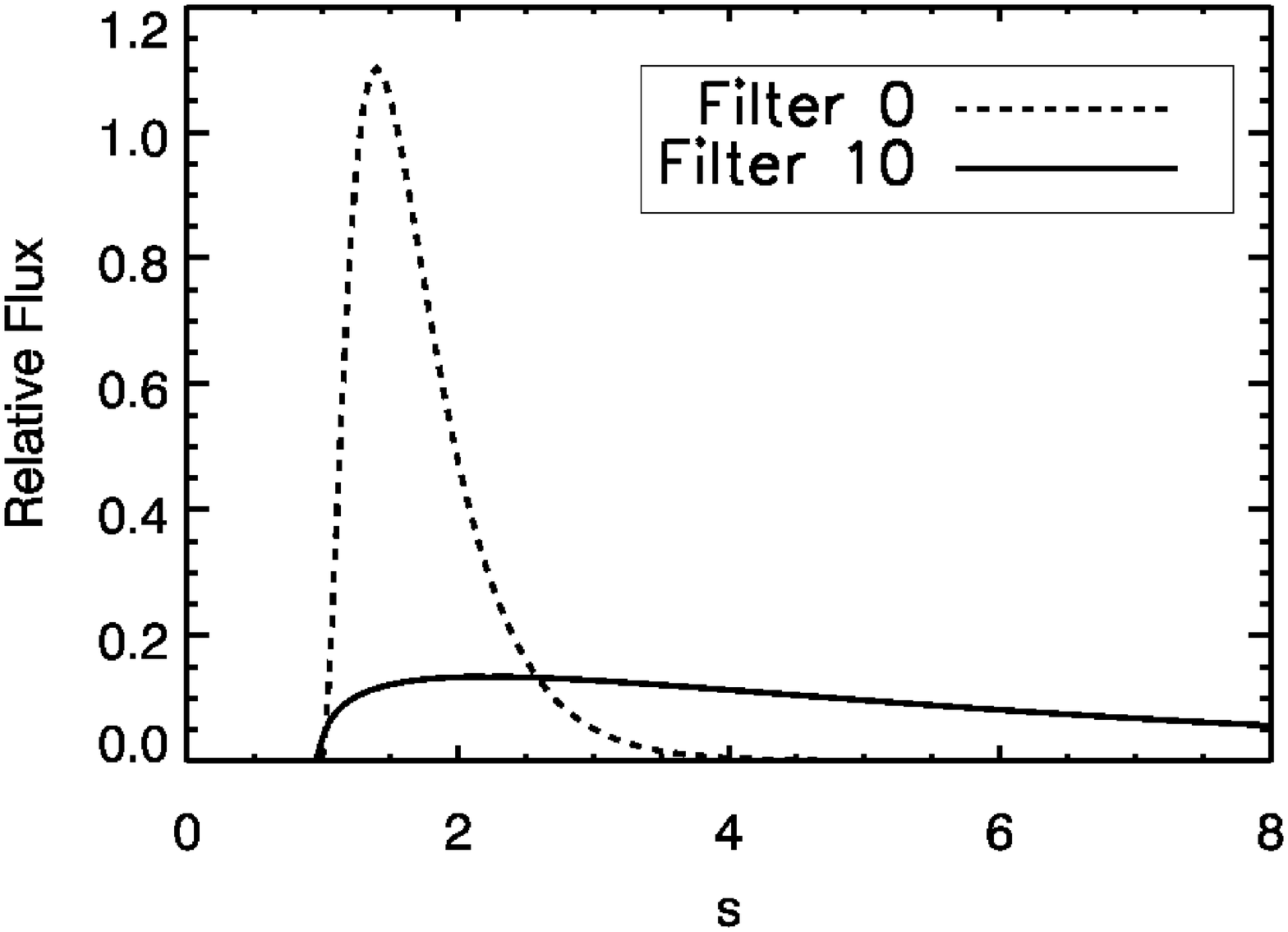}

\plotone{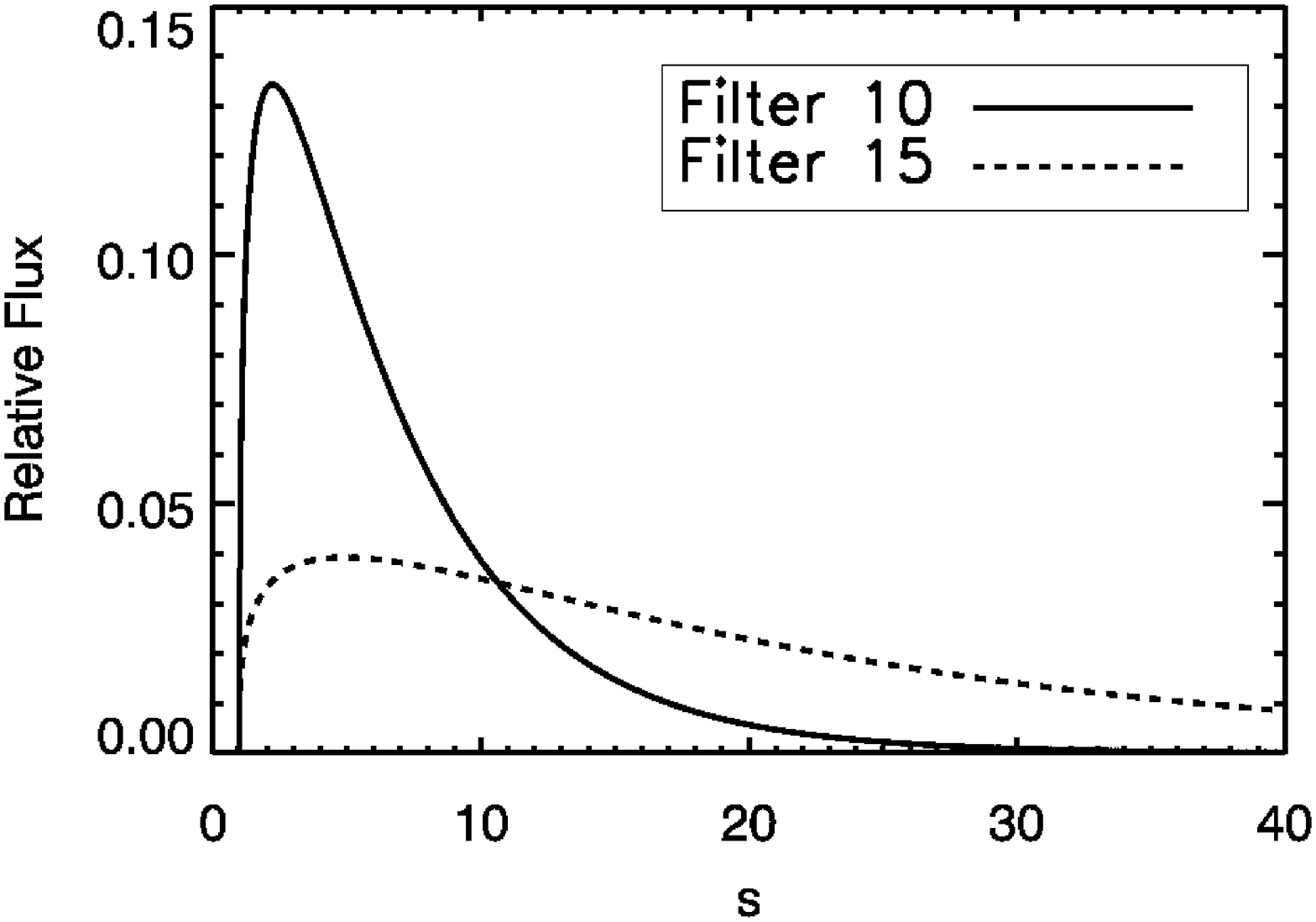}
\caption{Radial surface brightness distributions ($2\pi s B_i(s)$) for an $r_{in}=0.2 r_E$
Shakura-Sunyaev disk model in four filters, with
central dimensionless wavelengths $x_{-10}=0.0271$, 
$x_0=0.2014$, $x_{10}=1.498$, and $x_{15}=4.086$.  The vertical axis is normalized
so that the integrated surface brightness equals unity.
 \label{disks}}
\end{figure}

To create a model of the disk as it would be seen through a particular filter $
i$, we integrate the monochromatic specific intensity
over the wavelengths included in the filter:\footnote{For narrow
filters, the wavelength across a single filter can be
treated as a constant, $x_i$, as in~\citet{k03}.
This eliminates the need to do the integral in Equation~(\ref{ui}),
since $B_i(s)\approx B_{x_i}(s)$.  Note, however, that in~\citet{k03},
the factor of $(1-\sqrt{r_{in}/r})^{1/4}$ in Equation~(\ref{temp}) is
neglected, so those disk models differ significantly from ours for
$r\sim r_{in}$.}
\begin{equation}
B_i(s) = \int_{x_{i,min}}^{x_{i,max}} B_x(s) dx. \label{ui}
\end{equation}
This function $B_i(s)$, with units of erg~s$^{-1}$~cm$^{-2}$~sr$^{-1}$,
serves the same purpose for the Shakura-Sunyaev model as $G(r)$ and $C(r)$ do
for the Gaussian and cone models in the
previous sections, except that we use a dimensionless radius as the independent
variable and there is a different function for each filter.  We can put
$B_i(s)$ in a form similar to the surface brightness profiles of the other
disk models if we define the total flux at Earth from the disk in filter~$i$,
\begin{equation}
F_i \equiv \frac{2\pi r_{in}^2}{D_S^2}\int_1^{\infty}B_i(s)s ds, \label{li}
\end{equation}
and the normalized surface brightness,
\begin{equation}
b_i(s) \equiv \frac{B_i(s)}{2\pi\int_1^{\infty}B_i(s)s ds}. \label{reff}
\end{equation}
Then we can write the Shakura-Sunyaev disk surface brightness as
\begin{equation}
B_i(s)=F_i\frac{D_S^2}{r_{in}^2}b_i(s).
\end{equation}
Radial surface brightness profiles in four filters are shown in 
Figure~\ref{disks}.

The Shakura-Sunyaev disk model that we end up with depends on two
parameters: $r_{in}$, the innermost radius of the disk,
and $i$, the filter number.  The temperature $T_0$ only determines the relation
 between $\lambda$ and $x$.

\subsection{Other Models}

Our Shakura-Sunyaev disk model is similar to the thin accretion disk models 
used by~\citet{ak99} and~\citet{jwp92}.
Those models are more complicated,
however, as they include rotating black holes, tilted disks, and
relativistic effects.
Microlensing simulations with nonthermal models have also been
considered~\citep{rb91}, but we do not include
such models in this study.

\section{MAGNIFICATION MAPS}

The effect of microlenses on the total macroimage flux may be represented by a
 magnification map in the source plane, where the value at each point of
the map is equal to the magnification of the source
 at that point, relative to the average macroimage 
magnification~\citep{k86,p86,w90,wps90}.
The microlensing light curve of a small, point-like source can be found by tracing a path 
across the
magnification map~\citep[e.g.,][]{p86,wps90,k03}.
  For an extended source, we must first convolve the source profile with 
the magnification
map to find the magnification due to microlensing at each location 
in the source plane~\citep[e.g.,][]{w02}.

\begin{figure}
\epsscale{0.6}
\plotone{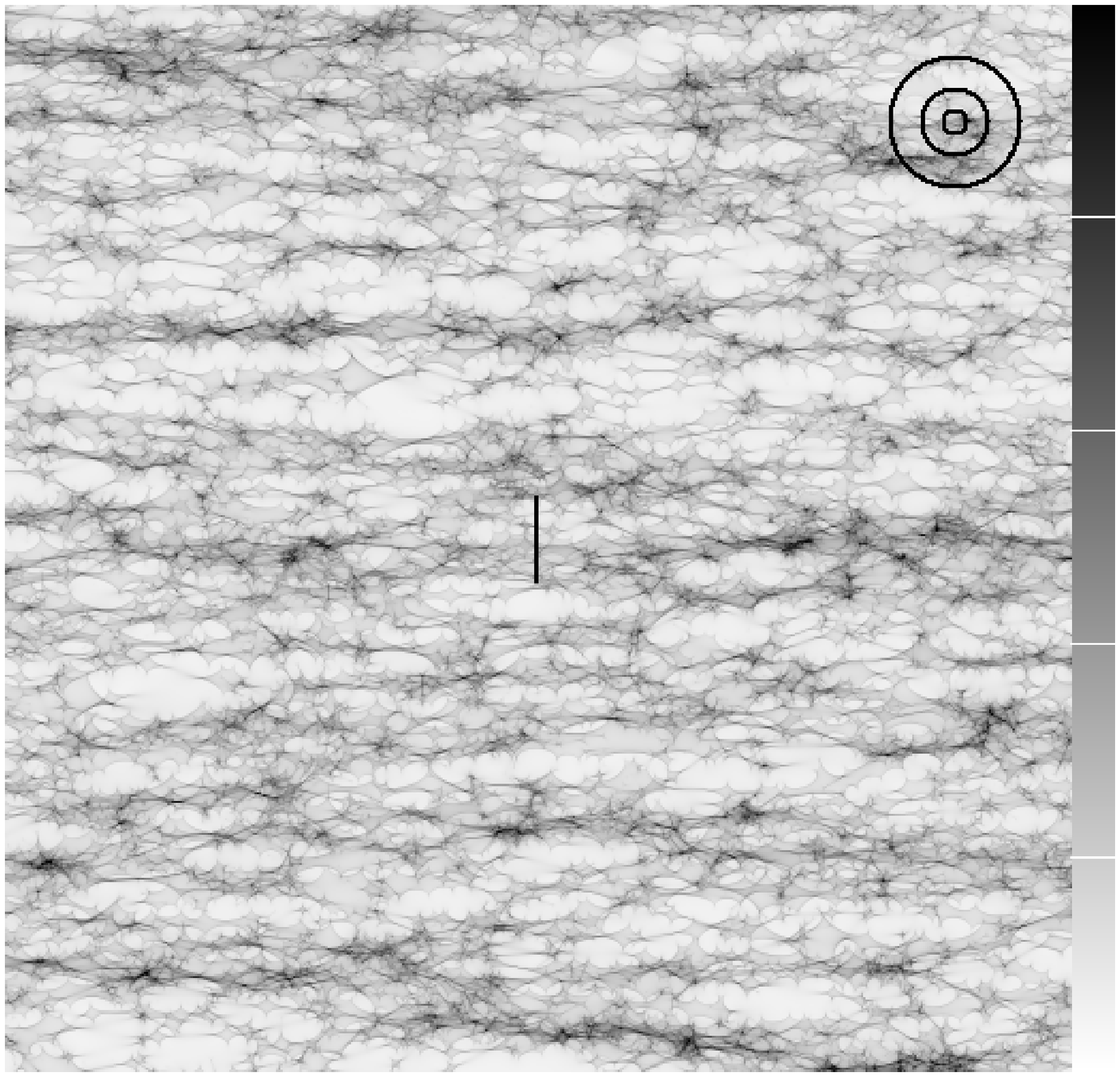}

\plotone{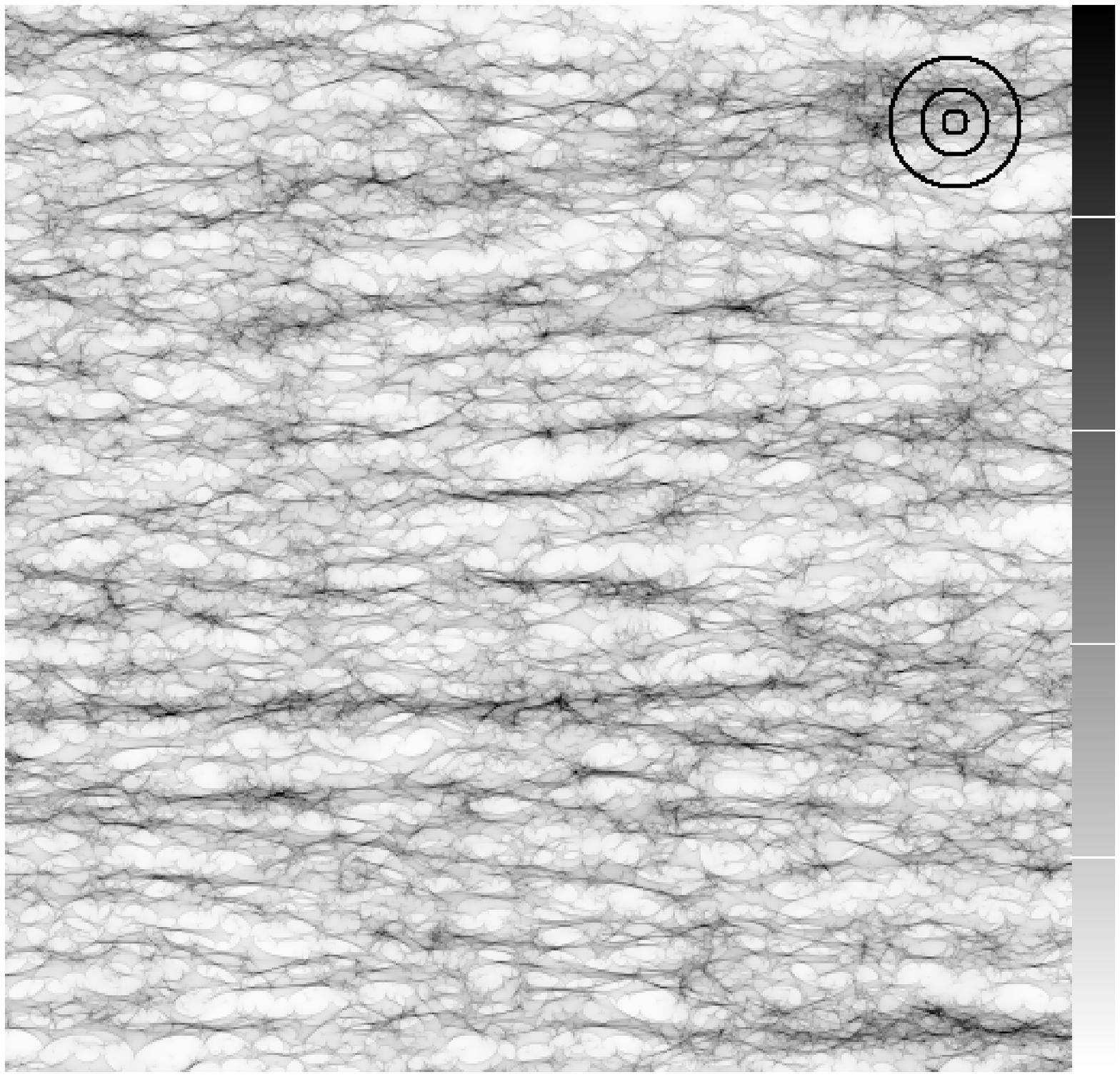}
\caption{Magnification maps in the source plane for a positive parity image with $\kappa=\gamma=0.4$ (top) and a negative parity image with $\kappa = \gamma = 0.6$ (bottom).  The length of each side is 100 Einstein radii.  The white lines on the greyscale bar correspond to magnifications
that are 1, 2, 3, and 4 times the average macroimage magnification.  Dark regions have greater magnification than light regions.  The black circles have radii of 1, 3, and 6 Einstein radii for comparison with the Shakura-Sunyaev disk models.  The black vertical line in the top map shows the path used for the light 
curves in Figure~\ref{lc}.\label{map}}
\end{figure}

The maps were made using ray-shooting techniques that simulate sending rays
from the observer through the
 lens to the source plane
\citep{k86,sw87,w90,wpk90,wps90,w99}.  The maps are 2000 by 2000
pixel arrays with sides of length 100
 Einstein radii.  We examined two cases typical of the images that
might be formed by a galaxy lensing a quasar: a positive parity image (minimum
of the time-delay
function) with convergence $\kappa = 0.4$
(all in compact objects),
 shear $\gamma = 0.4$, and theoretical average magnification $\mu = 5$; 
and a negative parity image
(saddle point) with $\kappa
 = 0.6$ (again, all in compact objects), $\gamma = 0.6$, and $\mu = -5$. 
Magnification maps for each case
are shown in Figure~\ref{map}.  The positive parity simulation included
37,469 lenses,
and the negative parity simulation included 56,224 lenses.

\begin{figure}
\plotone{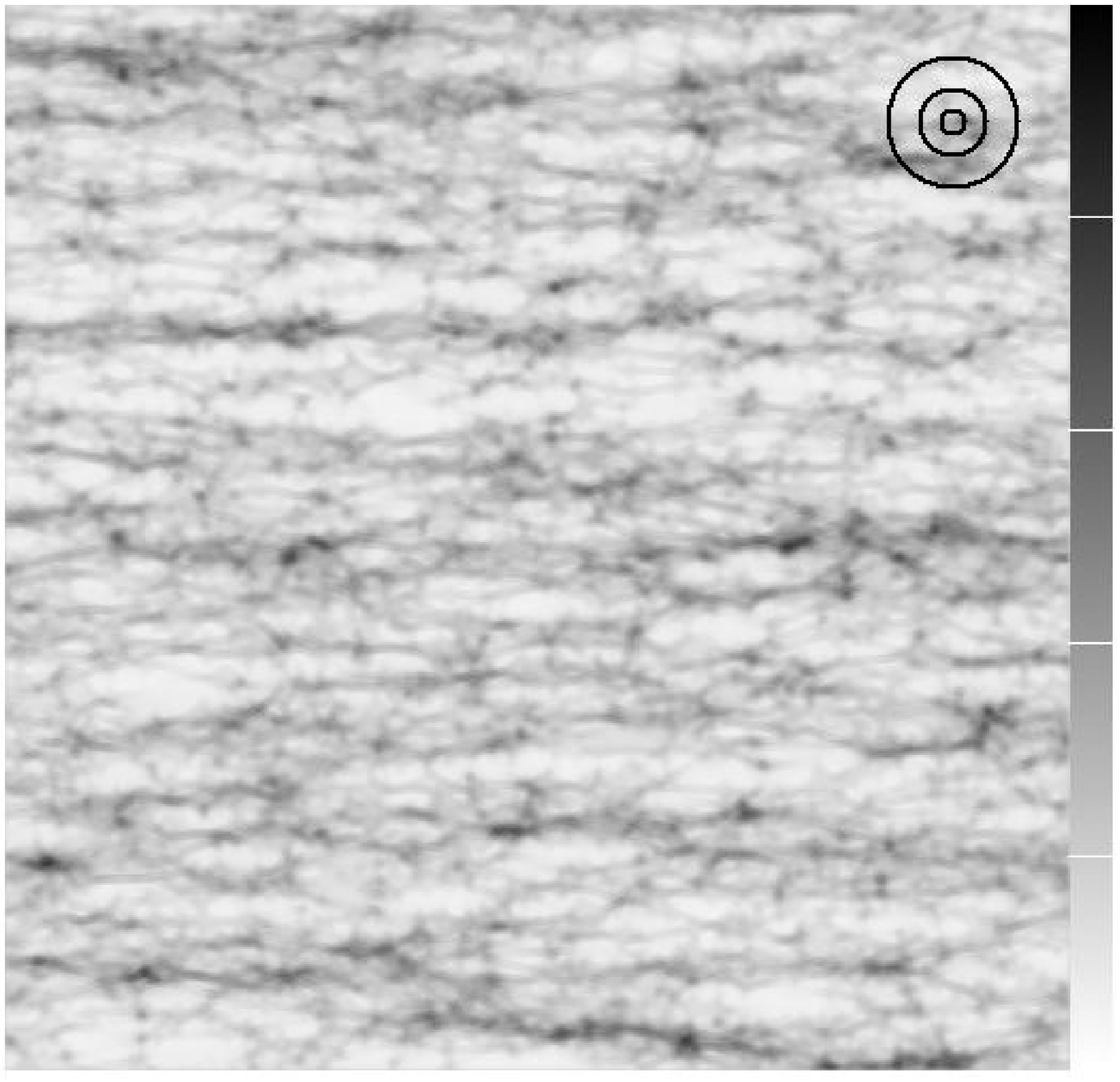}

\plotone{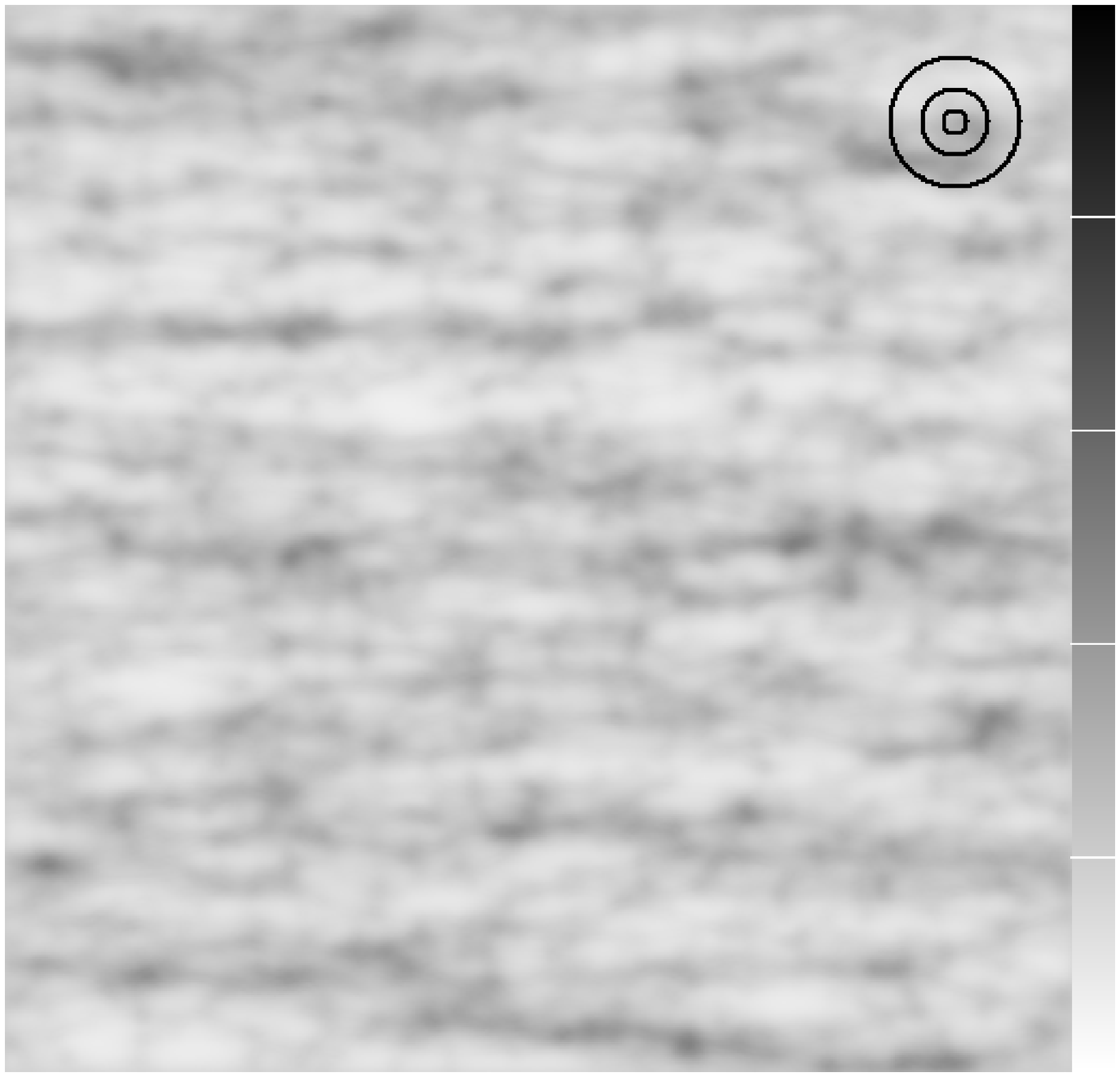}
\caption{Examples of magnification maps from convolving Shakura-Sunyaev disk profiles
with the original positive parity pattern in Figure~\ref{map}.
The innermost radius of each disk is $r_{in}=0.2r_E$.  For the top map,
the filter is $i=0$ with central wavelength $x_0$, the wavelength of the peak of the blackbody
distribution at the maximum temperature $T_0$; the disk surface brightness peaks
around $r=1.4r_{in}$ at this wavelength.  For the bottom  
map the filter is $i = 10$ with central wavelength $x_{10}=7.44x_{0}$, and
 the peak surface brightness is approximately at $r=2.2r_{in}$.
The scale and the reference circles are the same as in Figure~\ref{map}.\label{a20f0}}
\end{figure}

\begin{figure}
\epsscale{1.0}
\plotone{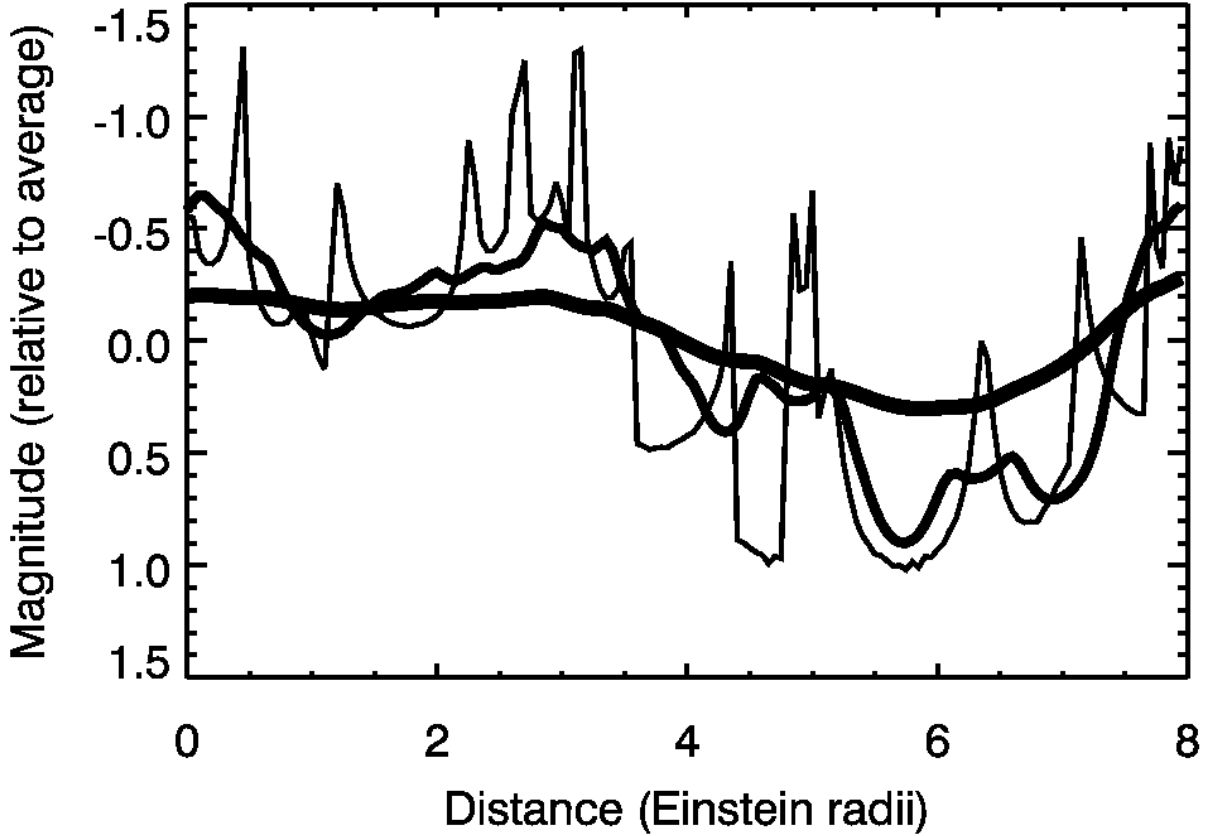}
\caption{Sample light curves from the positive parity magnification map in
Figure~\ref{map} and both maps in Figure~\ref{a20f0} ($\kappa=\gamma=0.4$).  The source
travels on a vertical path of length 4 Einstein radii in the center
of each map (the black line in the first panel of Figure~\ref{map}).  The thin curve is from the unconvolved positive parity map,
 the medium curve is from the convolution
with the disk viewed in the filter associated with the peak surface brightness at the
maximum temperature $T_0$ ($i=0$), and the thick curve is from the
convolution in the filter that is a factor of 7.44 longer in wavelength
($i=10$). \label{lc}}
\end{figure}

For each disk model we wished to study, we used the relevant equation from
Section~\ref{diskmod} to create a 2000 by 2000 pixel array for the 
disk brightness profile, $A$.  Let us call the original 
magnification map $M$.  By the convolution theorem, we can 
convolve $M$ and $A$ by
multiplying their two-dimensional Fourier transforms and then taking 
the inverse Fourier transform of the product.
This produces a new 2000 by 2000 pixel magnification map,
\begin{equation}
C = {\rm fft}^{-1}\left[{\rm fft}(M){\rm fft}(A)\right], \label{conv}
\end{equation}
where fft and fft$^{-1}$ stand for the fast Fourier transform and the
inverse fast Fourier transform, respectively~\citep[e.g.,][]{nr}.
Figure~\ref{a20f0} shows two examples of magnification maps from
convolutions with Shakura-Sunyaev disk models.  Sample light curves 
for paths through
these maps are shown in Figure~\ref{lc}.

The longest wavelengths used in our simulations were chosen so that at least 
95\% of the total 
accretion disk intensity would lie within the 2000 by 2000 pixel area of 
the magnification map.
 At longer wavelengths,
 the cooler temperatures of the disk at large radii make the outer 
regions of the disk more important than in the
shorter-wavelength filters.
If we use too long a wavelength, a large fraction of the disk intensity spills
out of
the area of our simulation, making the results inaccurate.
The wavelength at which this occurs varies with $r_{in}$.
  Although the cutoff is 95\%, for the majority of filters used the fraction of
 light included in the 2000 by 2000
 pixel area is above 99\%.

At the short-wavelength end, the cutoff was more arbitrary 
since the disk profiles and their magnification histograms
 do not vary much with wavelength beyond a certain point that depends
on the value of $r_{in}$.  We chose
to use wavelengths short enough to probe values of the
half-light radius (see Section~\ref{stats}) close to the inner radius
 $r_{in}$ (within one Einstein radius).

\section{MAGNIFICATION HISTOGRAMS} \label{histchap}

\subsection{Histograms of Convolutions with Shakura-Sunyaev Disks} 
\label{magsec1}

The values in a magnification map
 are ratios of the macroimage's flux at Earth when the source is at a particular point in
the map, $F({\emph{\bf r}})$, to the average macroimage flux, $\bar{F}=\mu F_s$, 
where $F_s$ is the flux of the unlensed source
at Earth.  We convert these ratios to magnitude differences,
\begin{equation}
\Delta m({\emph{\bf r}}) = -2.5 \log_{10} \left(\frac{F({\emph{\bf r}})}{\bar{F}}\right) 
\label{deltam}
\end{equation}
and plot a histogram of $\Delta m$ for the convolution with each disk model, as
 in~\citet{w92}.  The
number of pixels that fall into each bin of $\Delta m$ is represented 
as a probability for the macroimage to have a
certain magnitude shift by dividing the number of pixels in the bin 
by the total number of pixels.
Histograms of the original magnification maps are shown in 
Figure~\ref{historig}. Both histograms have two main peaks, typical for 
images with $|\mu|=5$~\citep{sw02}.
A minimum (positive parity) must have at least unit magnification,
so the positive parity histogram is cut off at the low-magnification end.  
At lower magnification,
the negative parity histogram has a tail that extends down to $\Delta m \sim 2-
3$~mag.  The left peak of each histogram (around
$\Delta m=1-1.5$~mag) is associated with the case of no extra microimage
minima, while the right peak around $\Delta m=0$~mag is associated 
with the case of one extra microimage pair~\citep{r92}.

\begin{figure}
\plottwo{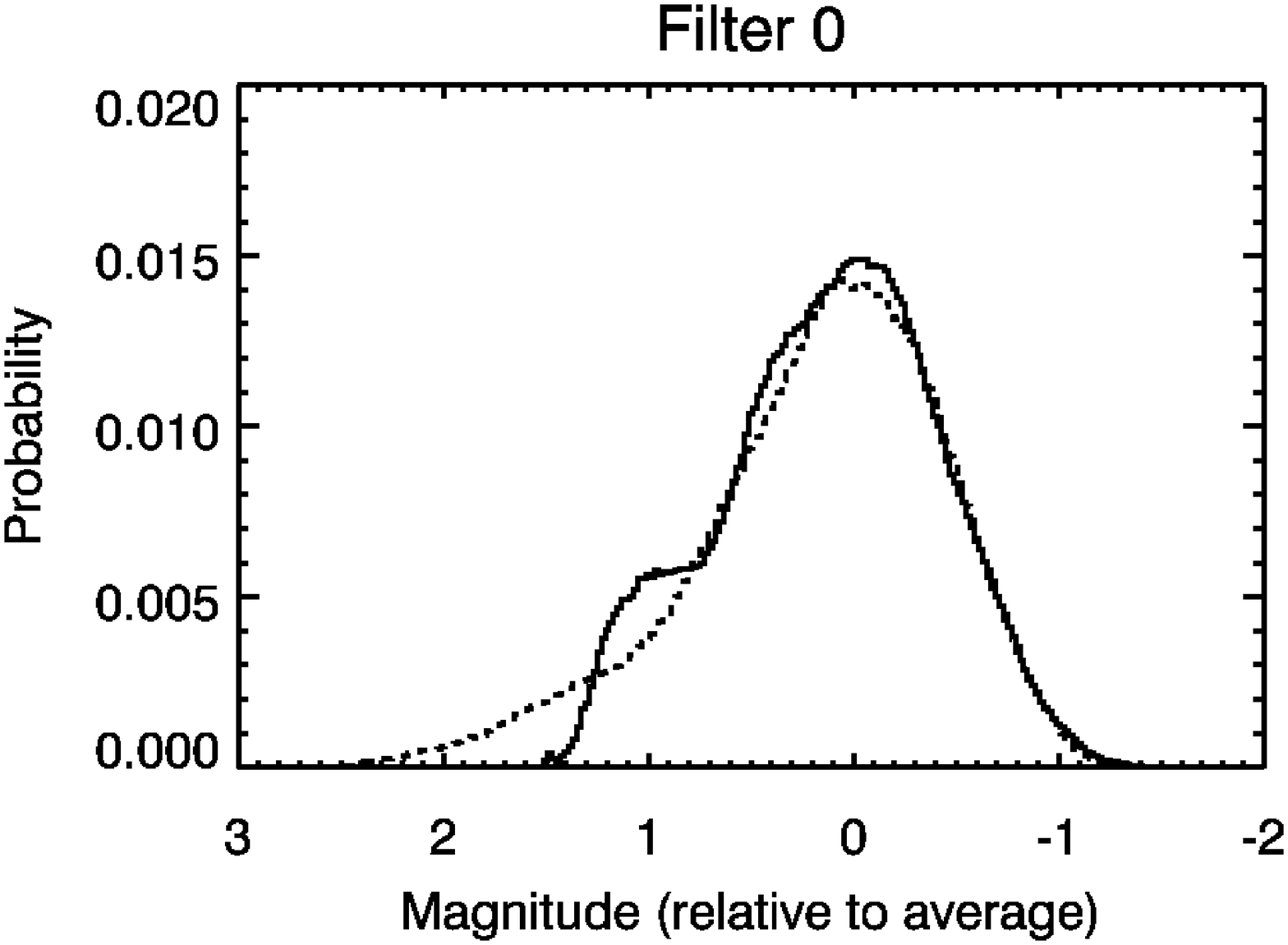}{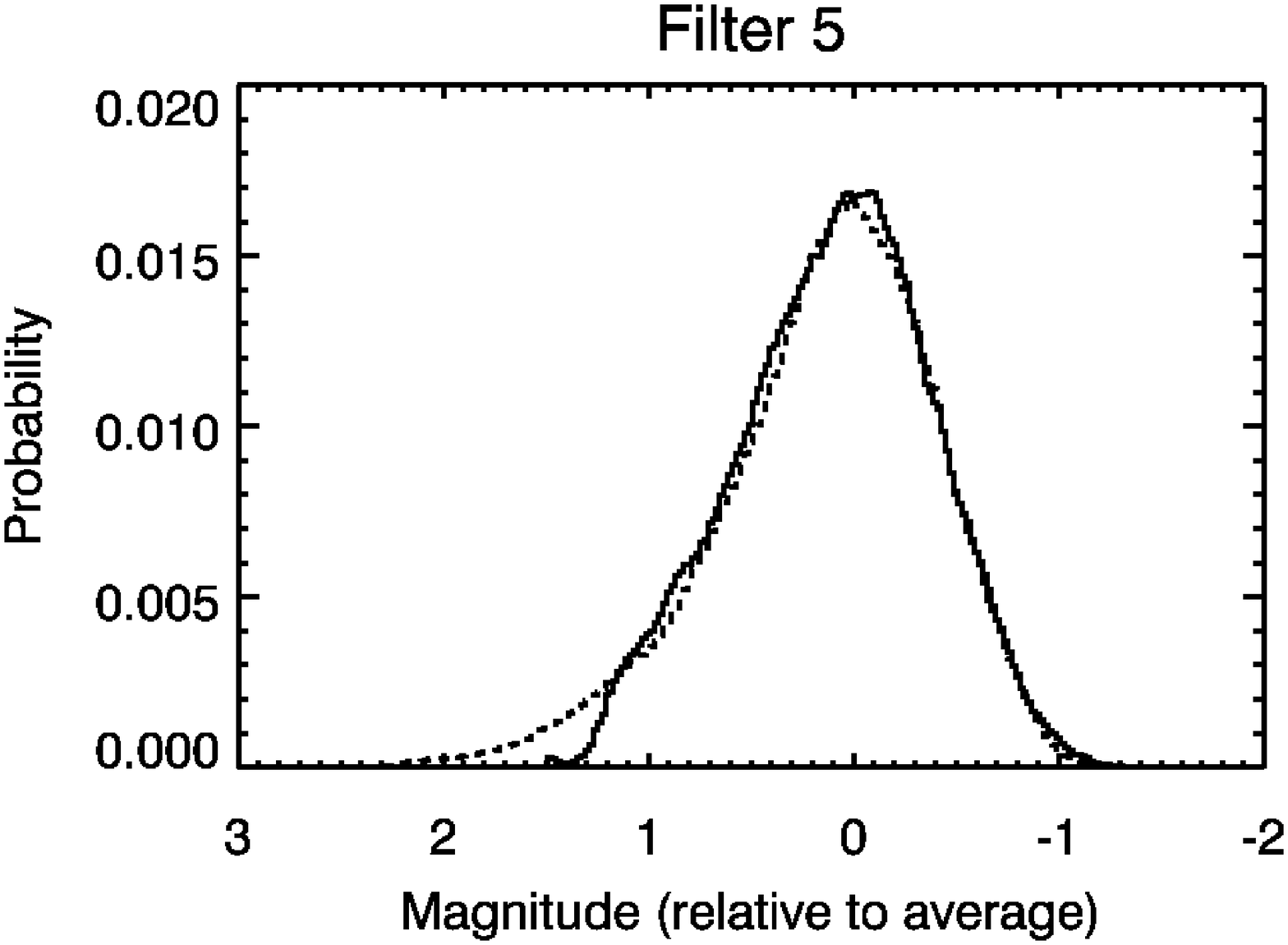}

\plottwo{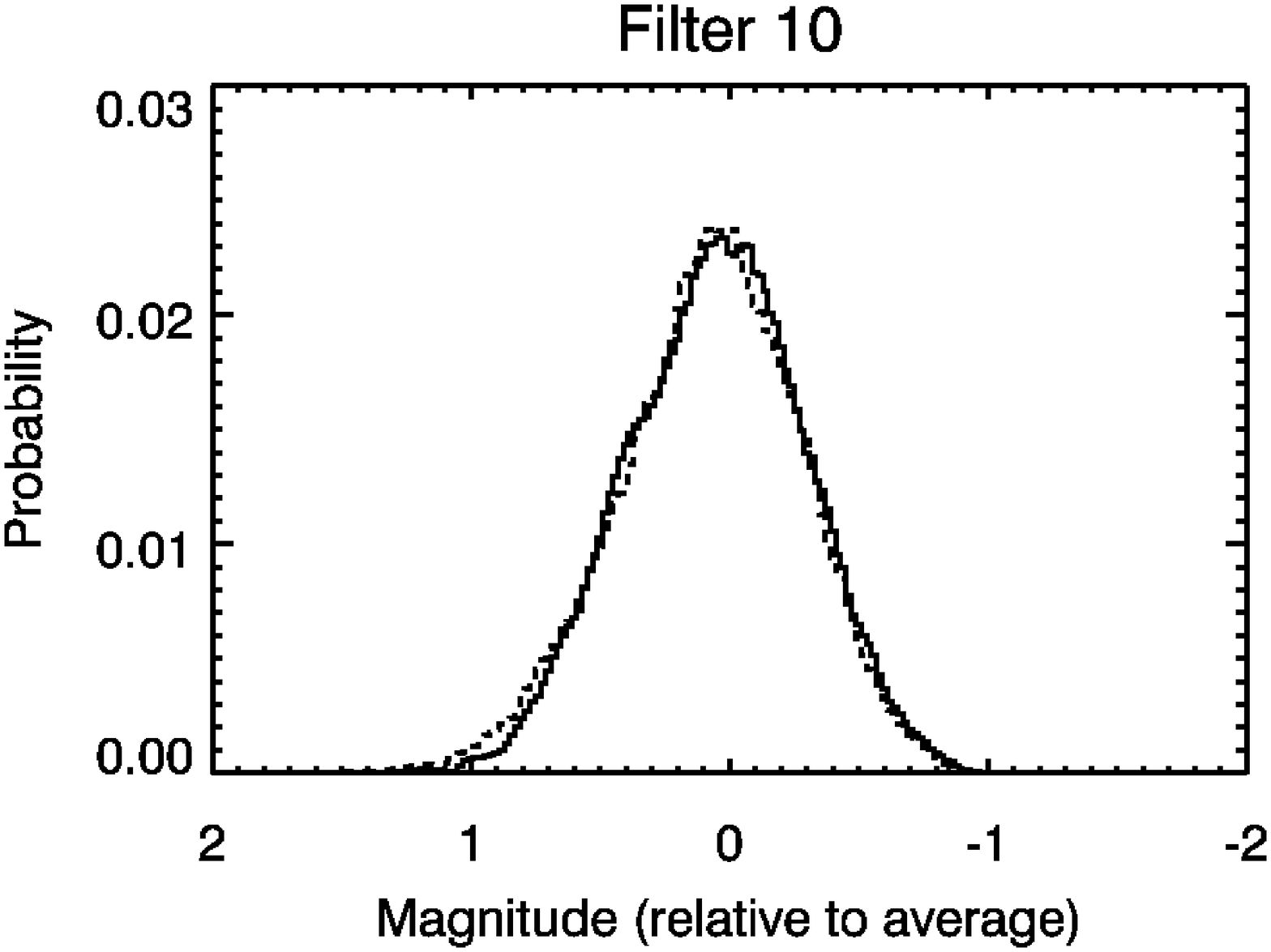}{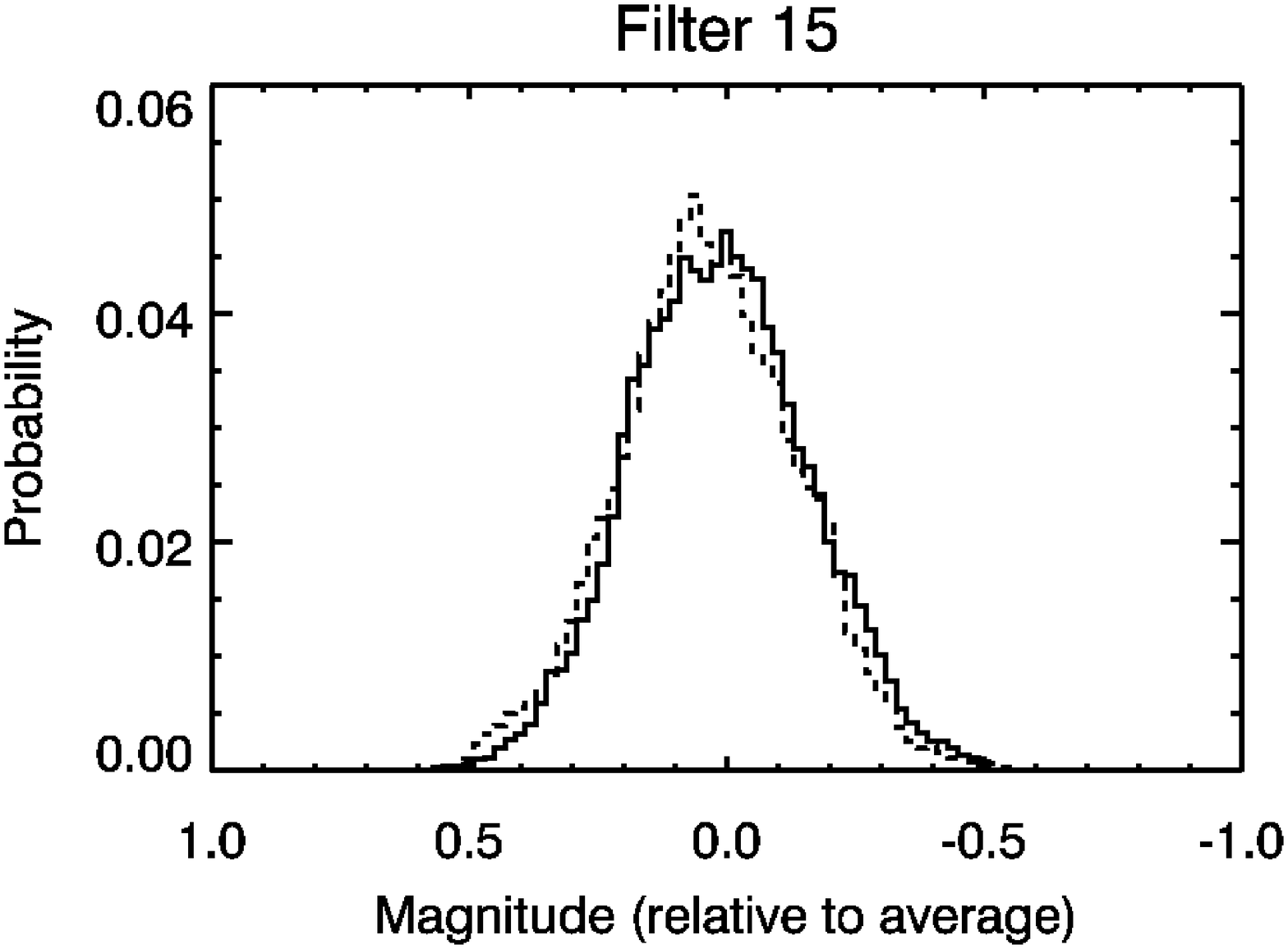}
\caption{Histograms of magnitudes (relative to the magnitude that corresponds
to the average macroimage flux at Earth)
for convolutions of Shakura-Sunyaev disk profiles with $r_{in}=0.2r_E$ in various filters
with the positive parity $\kappa = \gamma = 0.4$ magnification map (solid curves)
and the negative parity $\kappa = \gamma = 0.6$ magnification map (dashed curves).
The half-light radii of the disks used as sources are $0.28r_E$, $0.41r_E$,
$1.00r_E$, and $3.32r_E$, respectively. The histograms at shorter wavelengths
than that of the filter associated with the peak surface brightness at the
maximum temperature $T_0$ (upper left) are all very similar,  
so they are not shown here.  \label{hist1}}
\end{figure}

\begin{figure}
\plottwo{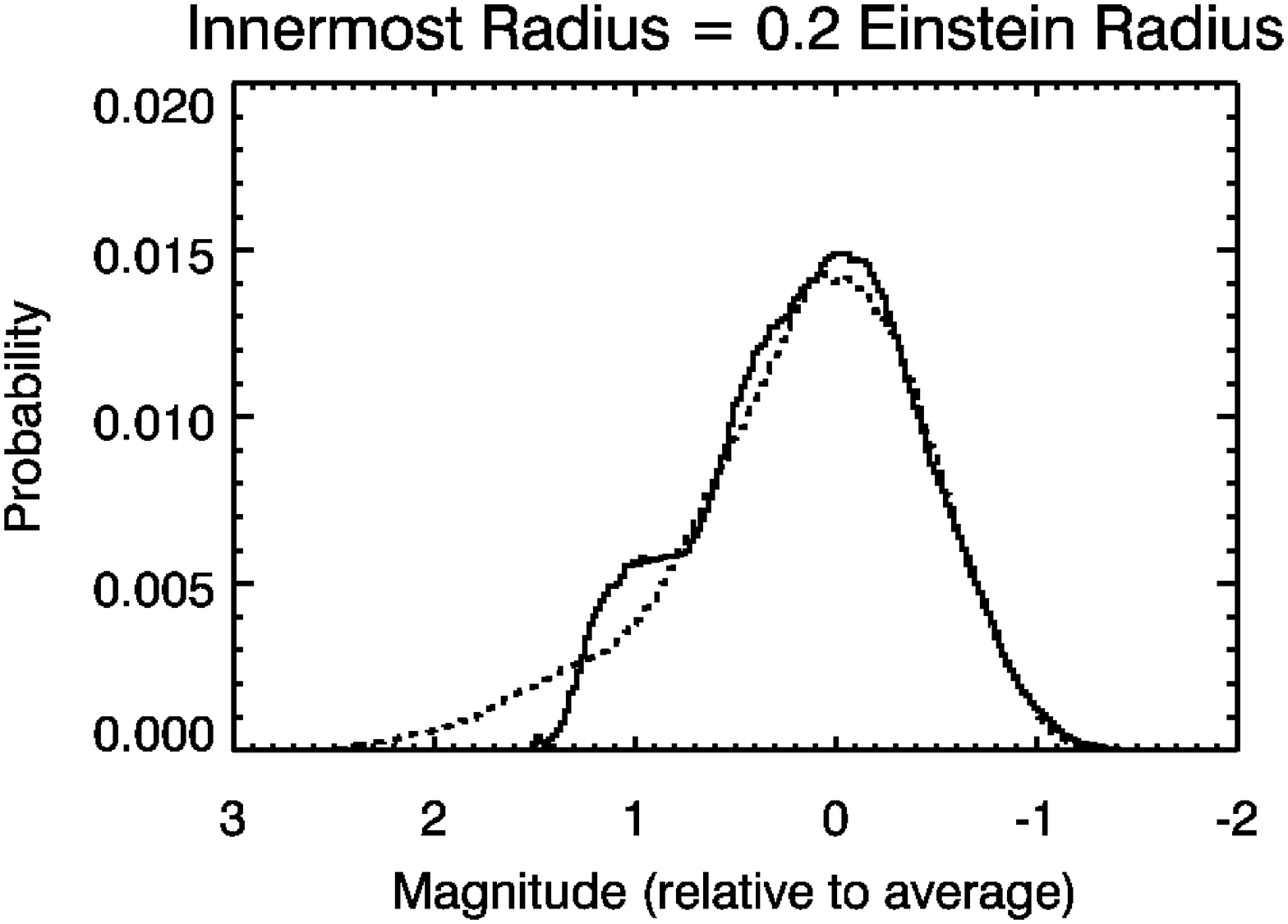}{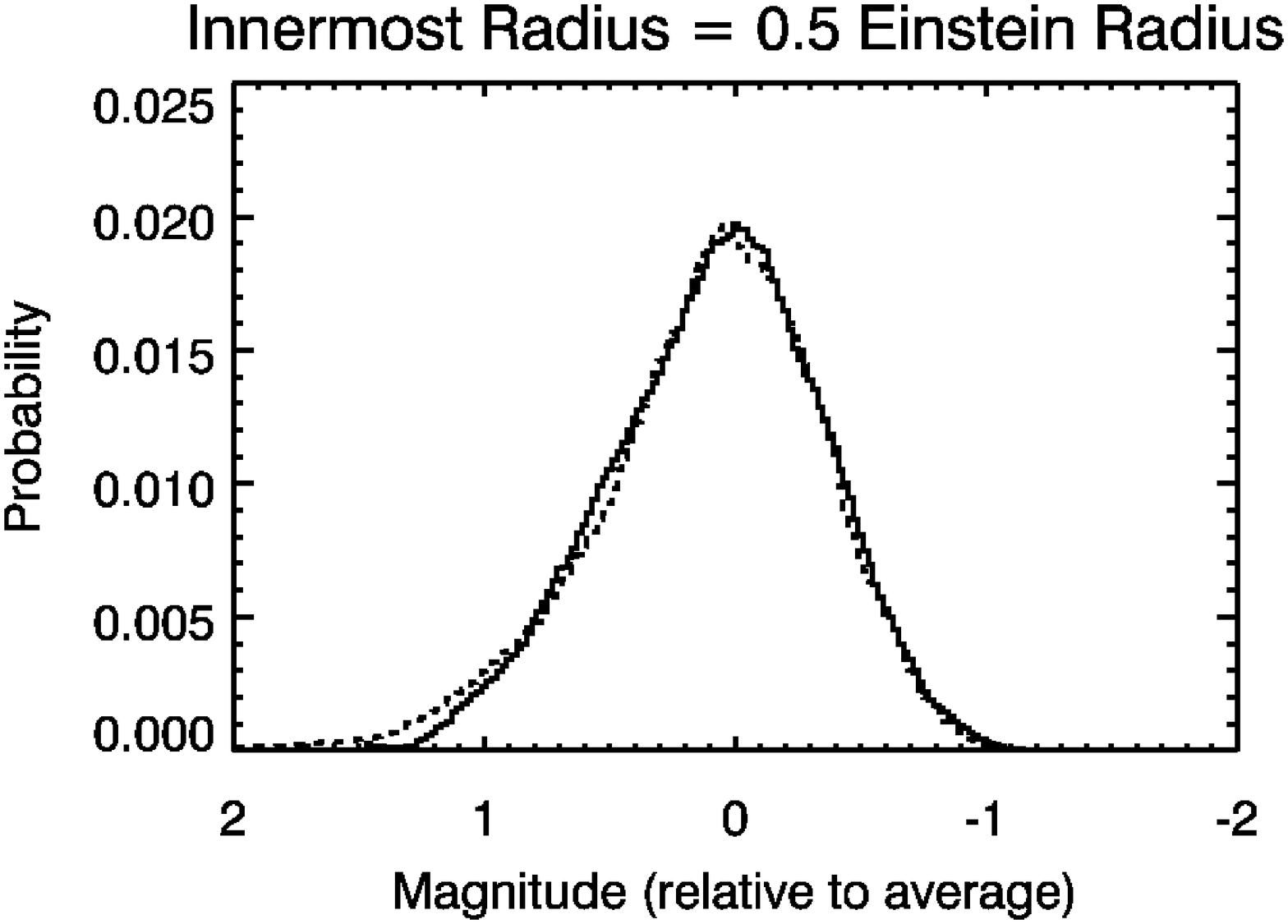}

\plottwo{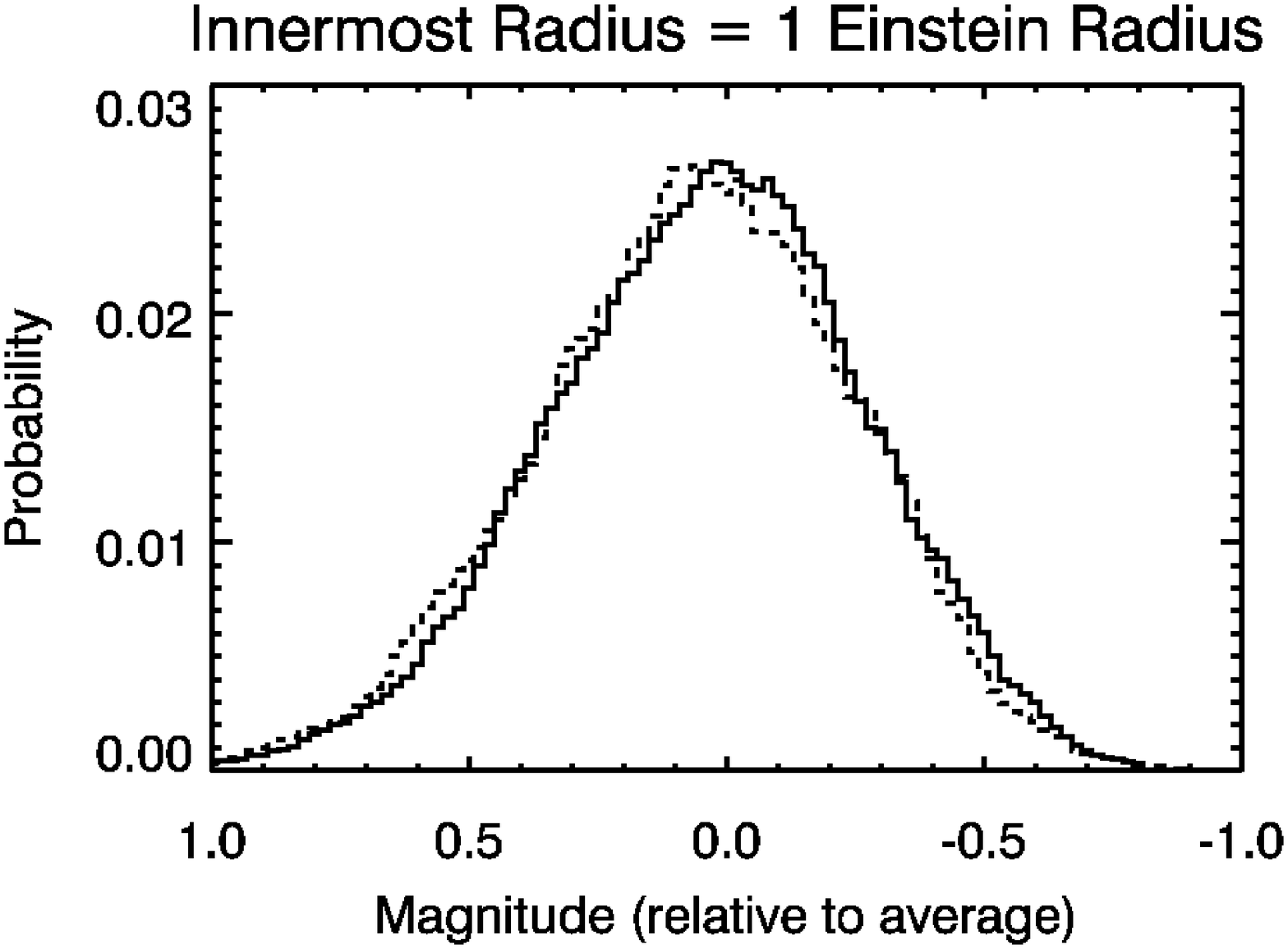}{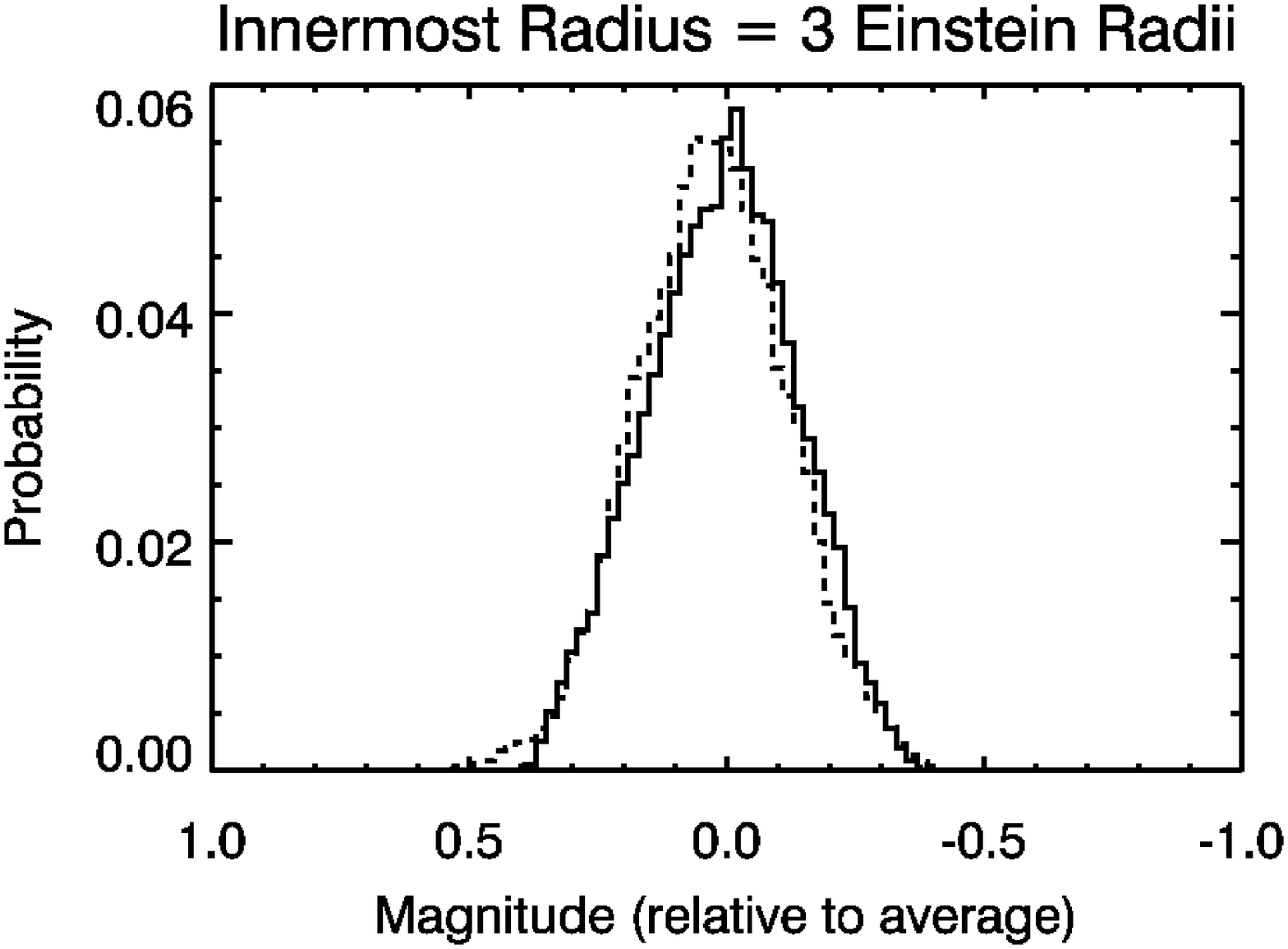}
\caption{Histograms of magnitudes relative to the average for convolutions of Shakura-Sunyaev disk profiles of various sizes in the filter associated with the peak
surface brightness at the maximum temperature $T_0$ ($i=0$)
 with the positive parity $\kappa = \gamma = 0.4$ magnification map (solid curves) and the negative parity $\kappa = \gamma = 0.6$ magnification map (dashed curves).
The half-light radii of the disks used as sources are $0.28r_E$, $0.77r_E$, $1.58r_E$, and $4.84r_E$, respectively. \label{hist2}}
\end{figure}

We constructed magnification maps for convolutions with Shakura-Sunyaev disks 
in several filters with $r_{in}=0.2r_E$, $0.5r_E$, $r_E$, and $3r_E$, where 
$r_E$ is the microlens Einstein radius.
  Histograms from some of these maps are shown in Figures~\ref{hist1} 
and~\ref{hist2}. 
For long wavelengths or large $r_{in}$, the histograms are sharply 
peaked at the
average macroimage magnification, and there is little difference 
between the positive and negative parity cases.
 These characteristics reflect the loss of detail in the magnification maps for
 convolutions with disks that have large effective sizes. 
As we discuss in Sections~\ref{typq} and~\ref{conc}, these disks are probably 
unrealistically large relative to the microlens Einstein radius, so the results 
that follow are valid in the limit of very small microlenses or very large 
disks.  However, some of the results we find should be true for more general 
lens systems.

\subsection{Histogram Statistics}\label{stats}

\begin{figure}
\plotone{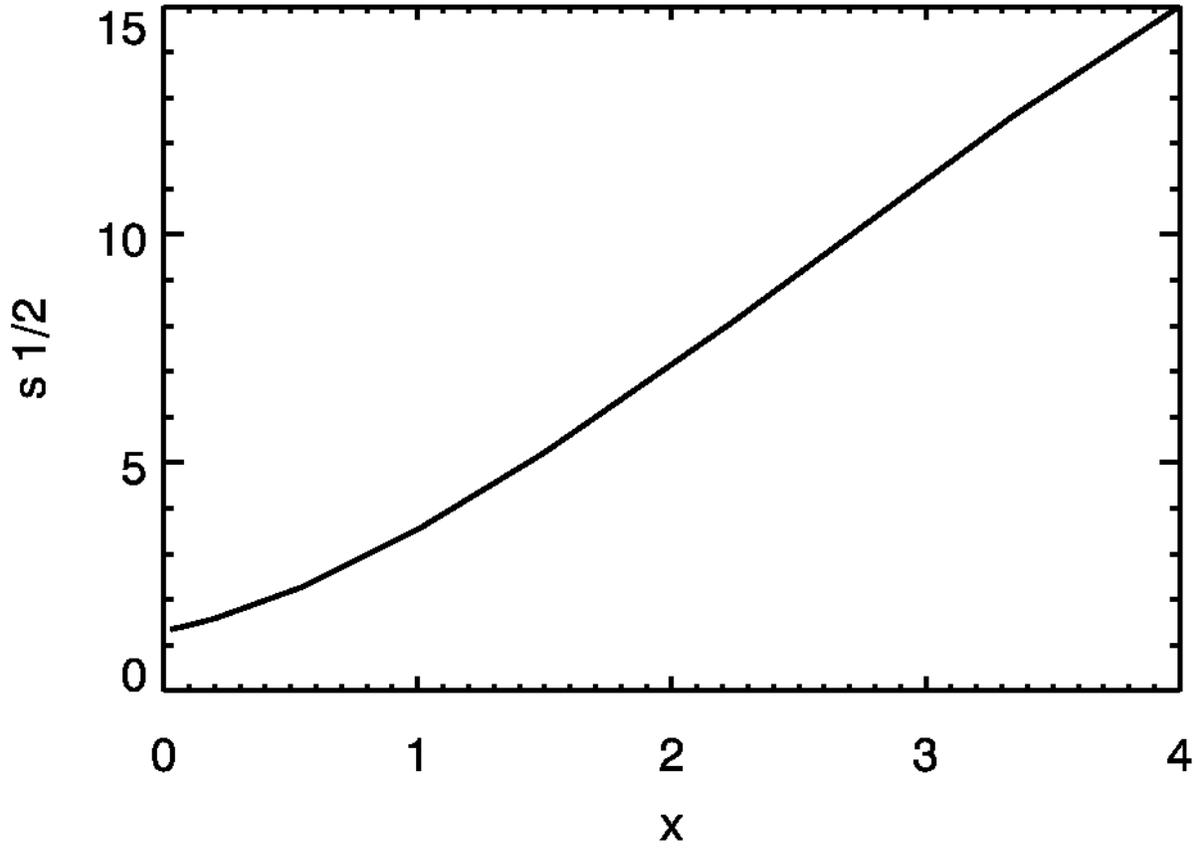}
\caption{Dimensionless half-light radius ($s_{1/2}=r_{1/2}/r_{in}$) versus dimensionless wavelength,~$x$, for the Shakura-Sunyaev models.\label{rx}}
\end{figure}

Since the surface brightness from disks in different filters falls 
off with radius at different rates,
we can use the half-light radius, $r_{1/2}$, as a proxy for wavelength (see Figure~\ref{rx}).
  For each magnification histogram, we calculated the dispersion
(root mean square or rms) and skewness of the data and plotted these
 statistics against $r_{1/2}/r_E$.  The results are shown in 
Figures~\ref{stat2} and~\ref{stat3}.

For all disk sizes, the dispersion decreases with $r_{1/2}/r_E$.
  This shows that the effect of microlensing is diminished at longer
wavelengths and for larger disks.  These trends are expected since the source
must be smaller than the microlens Einstein
radius for microlensing to play a significant role.

\begin{figure}
\epsscale{1.0}
\plotone{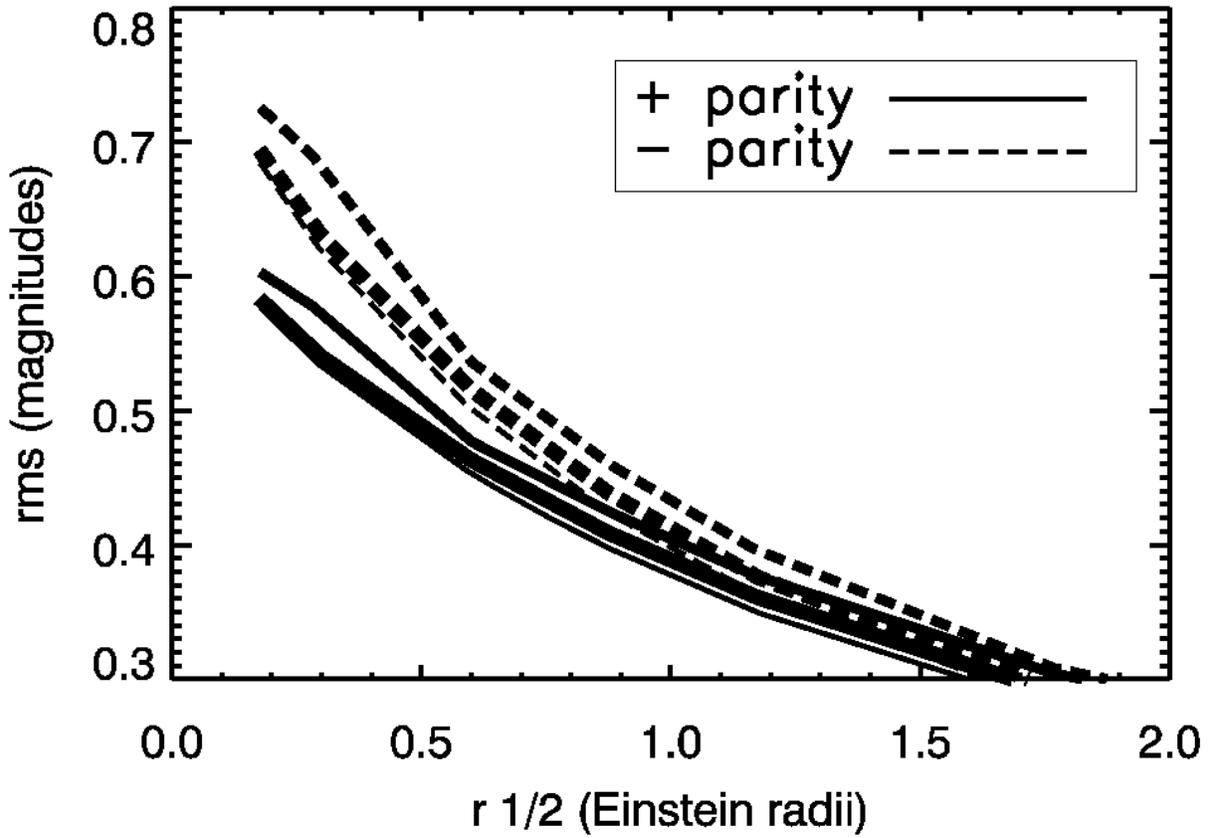}
\caption{Dispersion (rms) of histograms from convolutions of both positive 
($\kappa=\gamma=0.4$, solid curves) and  
negative ($\kappa=\gamma=0.6$, dashed curves) parity
magnification maps with Gaussian disks (thin curves), uniform disks (medium), 
and cones (thick).  For values of $r_{1/2}$ greater than about 2 Einstein radii,
the six curves shown here are nearly indistinguishable.\label{gdn}}
\end{figure}

Using the same methods described in Section~\ref{magsec1}, 
we produced magnification
histograms from convolutions with Gaussian disks, uniform disks, and cones.
These histograms
all have very similar
dispersion and skewness as a function of $r_{1/2}$; the dispersion results are
shown in Figure~\ref{gdn}.
From Figures~\ref{stat2} and~\ref{stat3} we see that, 
for a given value of $r_{1/2}$, there is little practical difference between 
the dispersions of histograms
produced with the Gaussian disks and those produced with the 
Shakura-Sunyaev accretion disk models.
This suggests that,
 to a good approximation, the microlensing fluctuations only depend on
$r_{1/2}$, and the disk may be modeled with
any reasonable surface brightness profile.
 We examine this claim more
quantitatively in the last paragraph of this section.

In the third moment of the histograms, the skewness,
we begin to see some greater differences between the Shakura-Sunyaev
models and the Gaussian models (lower panels in 
Figures~\ref{stat2} and~\ref{stat3}).
However, since skewness is much more
difficult to measure with observations than dispersion,
these differences may well be unimportant for most applications.

We also used chi-square tests to compare histograms from convolutions with
disks that have different shapes and different sizes.  Histograms associated
with uniform disks require about 10,000 independent observations to distinguish
 them with 95\% confidence from histograms associated with Shakura-Sunyaev
disks of the same size; the comparisons between the Gaussian disks or cones
and the Shakura-Sunyaev disks need an even greater number of observations,
around 40,000.  In contrast, the size comparisons tend to require far fewer
observations.  After examining many sizes of Gaussian disk models and comparing
the histograms of their convolutions, we found that to tell apart histograms 
associated with Gaussian disks that
differ in size by $0.25r_E$ (a quarter of an Einstein radius), it requires
2000 to 4000 independent observations to reach 95\% confidence.  If we make the
difference in size much smaller, the number of observations can be as large
as for the shape comparisons (for example, a $0.05r_E$ size difference in
Gaussian disks calls for around 40,000 observations for 95\% confidence), but
in this case the disks with different sizes are intrinsically much more
similar than the disks with different shapes, so it is no surprise that
the histograms that arise from convolutions with the disks with slightly
different sizes are also very similar to each other.

\subsection{Physical Values for Typical Quasars} \label{typq}

The results of the previous sections for Shakura-Sunyaev disks 
are given in terms of a dimensionless 
wavelength, $x$, and radius, $r_{in}$.  To understand how these results might 
apply to actual microlensed quasars, we must convert this wavelength and 
radius into physical quantities.

For a ``typical'' quasar, we will assume that there is a central black hole with 
mass $M=10^8$~M$_{\odot}$ and that the bolometric luminosity of the quasar is 
$L=10^{46}$~erg~s$^{-1}$~\citep[e.g.,][]{fkr92}.  From \citet{yt02}, we will take
 the efficiency for the quasar to be $\eta=0.2$, which gives an accretion rate 
$\dot{M}=5\times 10^{26}$~g~s$^{-1}$.  Doing a simple Newtonian calculation with
these numbers yields an innermost radius $r_{in}=2.5 M=3\times 10^{14}$~cm.  These 
values of $\dot{M}$ and $r_{in}$ are close to the typical quasar values given in 
\citet{fkr92}.  Using the formulas for a Kerr black hole from \citet{bpt72}, we can 
quantify the error due to the Newtonian calculation.  An innermost stable circular 
orbit at $r_{in}=2.5 M$ corresponds to a black hole spin of $a=0.879$.  This gives 
a binding energy per mass of 0.146, which is reasonably close to the assumed value 
of $\eta=0.2$ at the level of accuracy at which we are working.

By comparing the constant factor in the temperature-radius relation found in 
\citet{fkr92} to that in Equation~(\ref{temp}), we find that the maximum disk 
temperature is
\begin{equation}
T_0=0.488\left(\frac{3GM\dot{M}}{8\pi \sigma r_{in}^3}\right)^{1/4},
\end{equation}
where $G$ is Newton's constant and $\sigma$ is the Stefan-Boltzmann constant.  
Using the values listed above for $M$, $\dot{M}$, and $r_{in}$, the maximum temperature 
is $T_0=7.4\times 10^4$~K.

Using these results, we can compare the filters of the Shakura-Sunyaev disk model to a
real filter.  For example, the Sloan $r'$ filter covers a range of wavelengths
from about $5560~{\rm \AA}$ to $6950~{\rm \AA}$ \citep{f96}.  Taking the maximum
temperature of the accretion disk to be $T_0=7.4\times 10^4$~K and assuming that
 the source is at redshift $z_S=2.0$, the Sloan $r'$ filter
corresponds to a range of dimensionless wavelengths $0.95 < x < 1.18$.  This
is closest to the filter we label $i=8$, which has a range $0.90<x<1.10$.
The filter $i=8$ falls in the middle of the range of filters used in this study, 
so the artificial filters we used are close approximations to some real filters.

\begin{figure}
\plotone{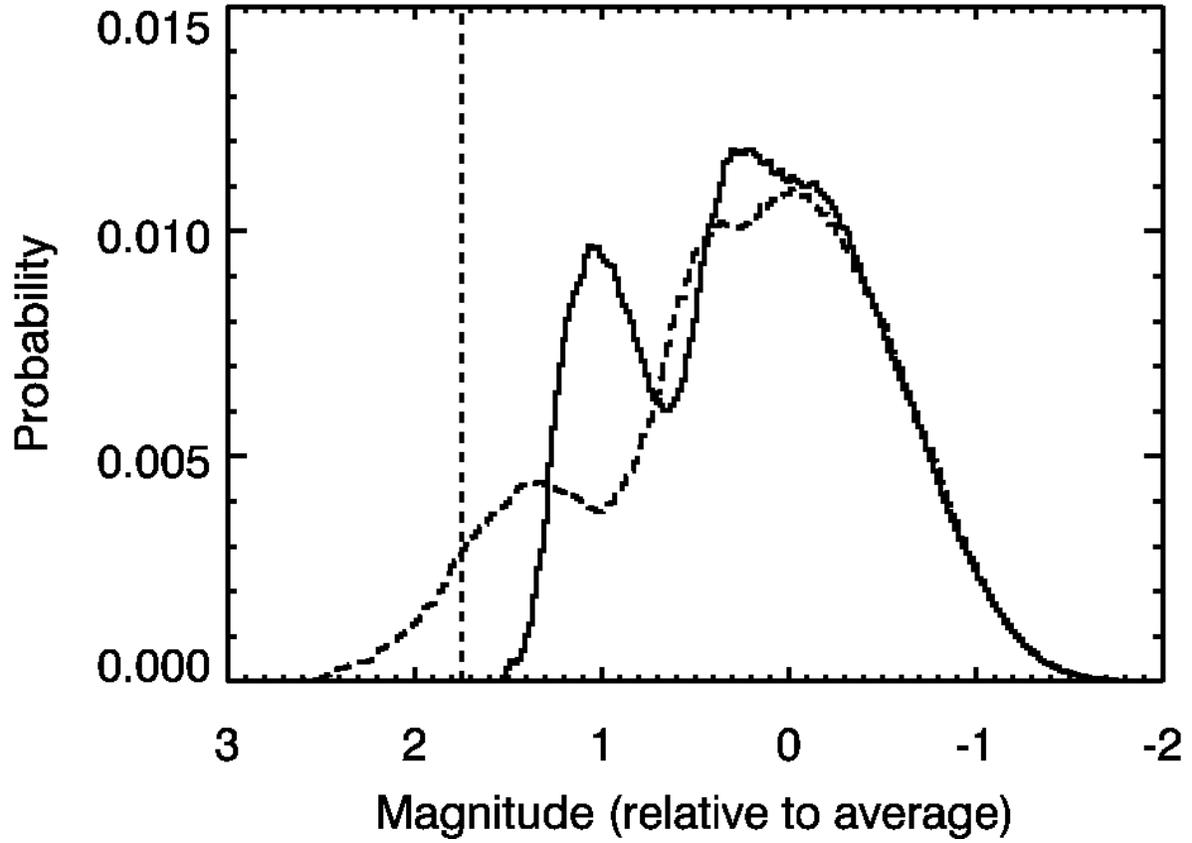}
\caption{Magnification histograms for the unconvolved magnification maps in
Figure~\ref{map}. The solid histogram is from the positive parity image and the dotted one is from the 
negative parity image. The bin width for each histogram is 0.02 mag.  The 
dispersions of the solid and dotted histograms are 0.67 and 0.81, respectively.
 The dashed vertical line shows the cutoff at unit magnification for the 
positive parity image.
\label{historig}}
\end{figure}

Next, we can compare the radii of the Shakura-Sunyaev disk models to physical 
radii. As mentioned earlier, our typical quasar has an innermost radius 
$r_{in}=3\times 10^{14}$~cm.  For a 
lens at redshift $z_L=0.5$ and a source at redshift $z_S=2.0$, the Einstein 
radius of a 1-M$_{\odot}$ microlens is $r_E\approx 5.7\times 
10^{17}$~cm~\citep{w92}.  
With these values, then, $r_{in}=0.0005r_E$.  This is considerably smaller 
than the $r_{in}$ to $r_E$ ratios examined in this study.  Of course, the 
exact ratio depends on the various masses and other parameters that we 
assume, but to have $r_{in}\sim r_E$ requires either very massive black holes 
or very small microlensing objects.  Therefore, 
Shakura-Sunyaev disks with physically realistic sizes are likely to be smaller
than the disks we modeled by at least an order of magnitude.  However, the
smallest disk we considered ($r_{in} = 0.2r_E$) produces magnification histograms
that are at least qualitatively similar, at short wavelengths, 
to the histograms for a point source
(Figure~\ref{historig}).  As we reduce the disk size from $r_{1/2}=0.28 r_E$ 
(the $r_{in}=0.2r_E$ disk in filter $i=0$) to $r_{1/2}=0$, the magnification 
histogram changes from that in the upper left panel of Figure~\ref{hist1} 
(dispersion 0.53 for positive parity, 0.62 for negative parity) to that in 
Figure~\ref{historig} (dispersion 0.63 for positive parity, 0.77 for negative 
parity).  A disk with a realistic half-light radius would have a magnification 
histogram that is practically identical to the corresponding histogram in 
Figure~\ref{historig}.   

This result suggests that, for ``typical'' Shakura-Sunyaev disks, not
only the shape but also the size can be ignored in most cases, so the source
behaves like a point source to a good approximation.  Therefore, we would not
expect to see significant chromatic effects for typical Shakura-Sunyaev disks.
An important exception is high-magnification events that occur when the source
crosses a caustic~\citep[e.g.,][]{wwt00,y01,sg02,w02,su03}, but away from
caustics the results found here apply.

\section{CONCLUSIONS} \label{conc}

We have produced several magnification histograms by convolving quasar source brightness
profiles with a variety of shapes and sizes with both positive and negative parity
image magnification patterns.  These histograms can be thought of as distributions of
the probability to observe the quasar macroimage with a certain magnification.  We compared
histograms associated with accretion disks of different shapes and different sizes by computing
moments of the histograms (dispersion and skewness), and by computing
chi-square values for pairs of histograms.

\begin{figure}
\epsscale{0.8}
\plotone{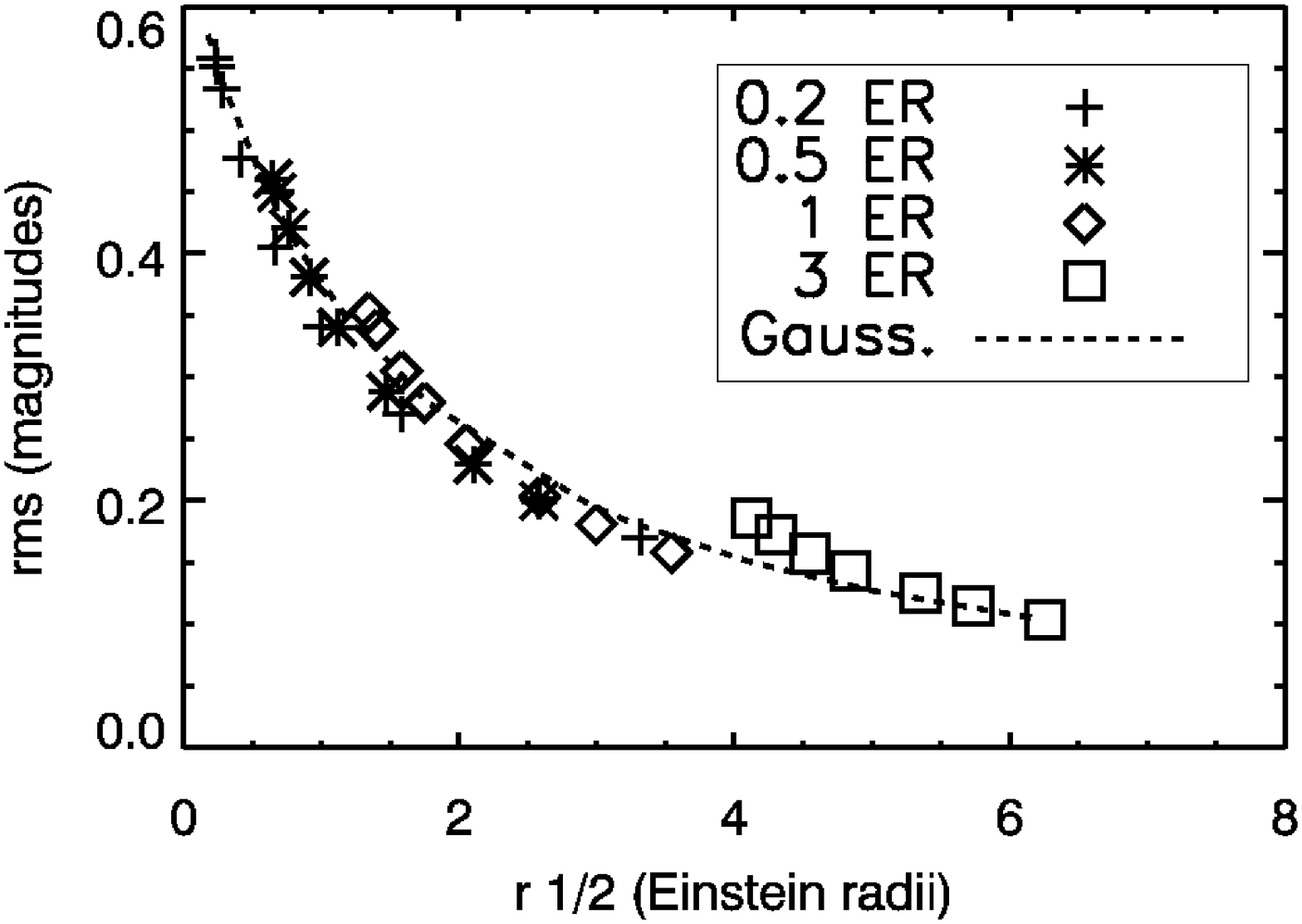}

\plotone{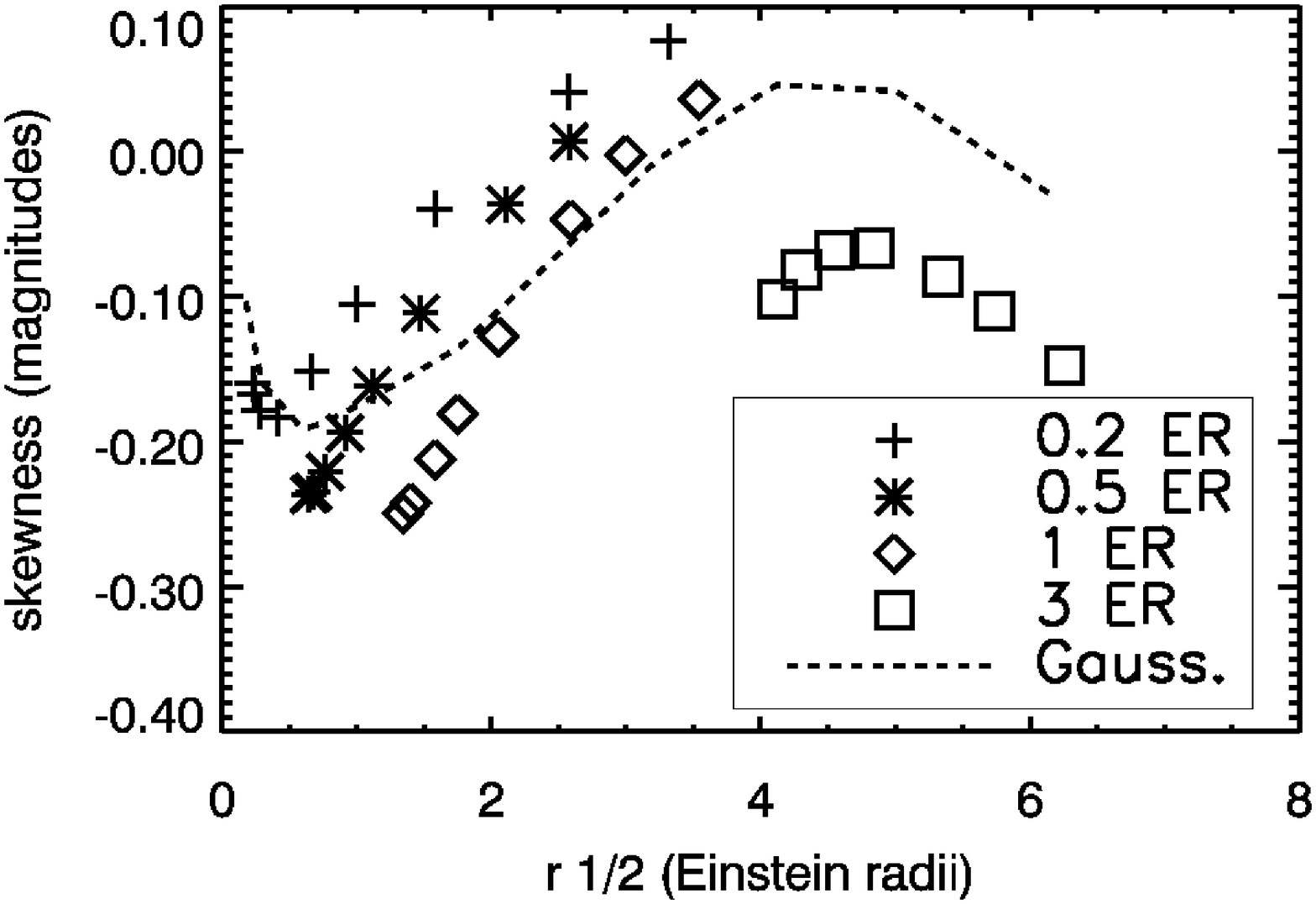}
\caption{Dispersion (rms) and skewness of convolutions of the $\kappa=\gamma=0.4$ magnification map with various Shakura-Sunyaev disk profiles.  Different plot symbols are used for
different values of $r_{in}$ (given in Einstein radii).  Dashed curves for the Gaussian disk models are shown for comparison.  Note that negative skewness is associated with a tail toward dimmer (positive) magnitudes.\label{stat2}}
\end{figure}

\begin{figure}
\plotone{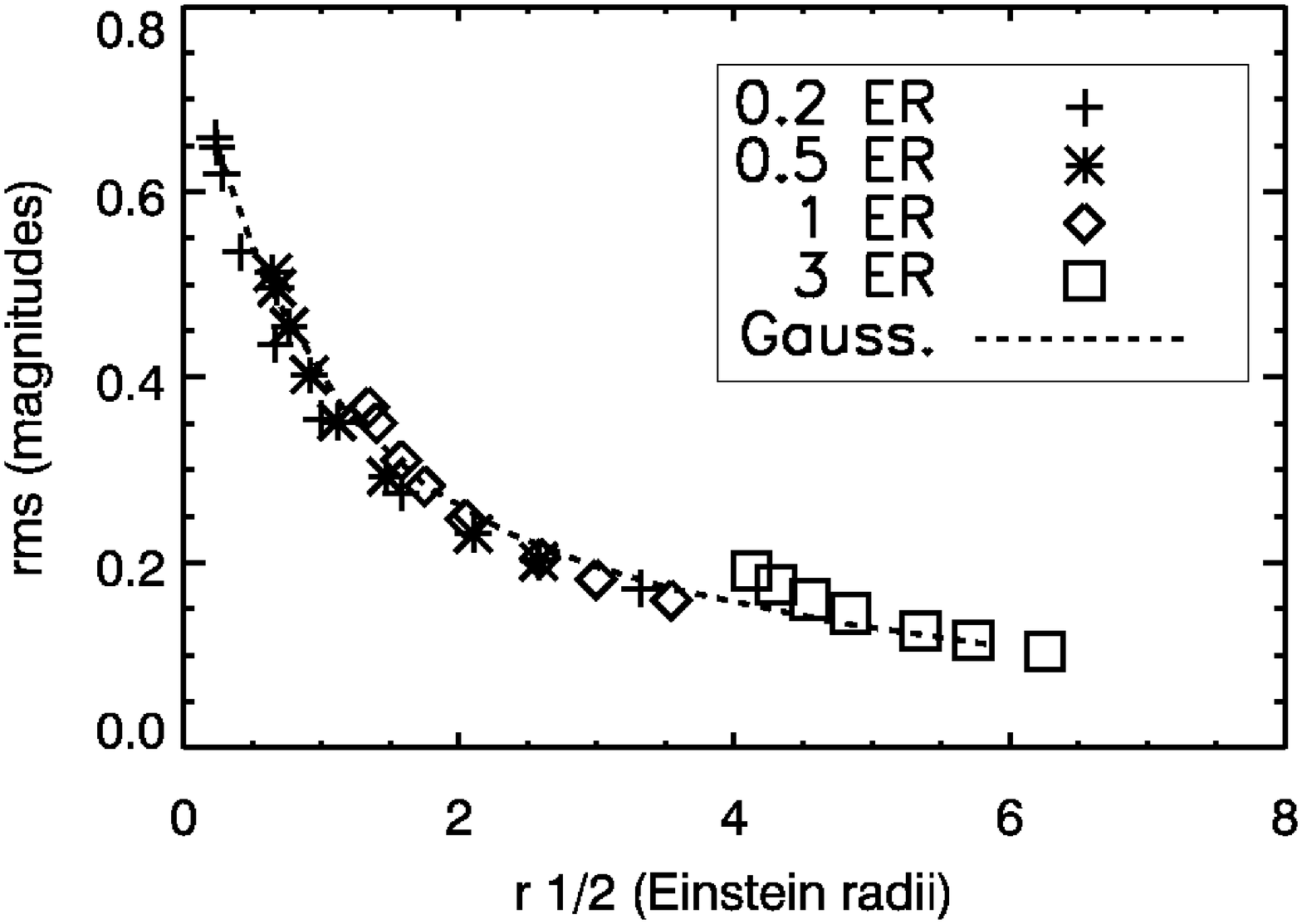}

\plotone{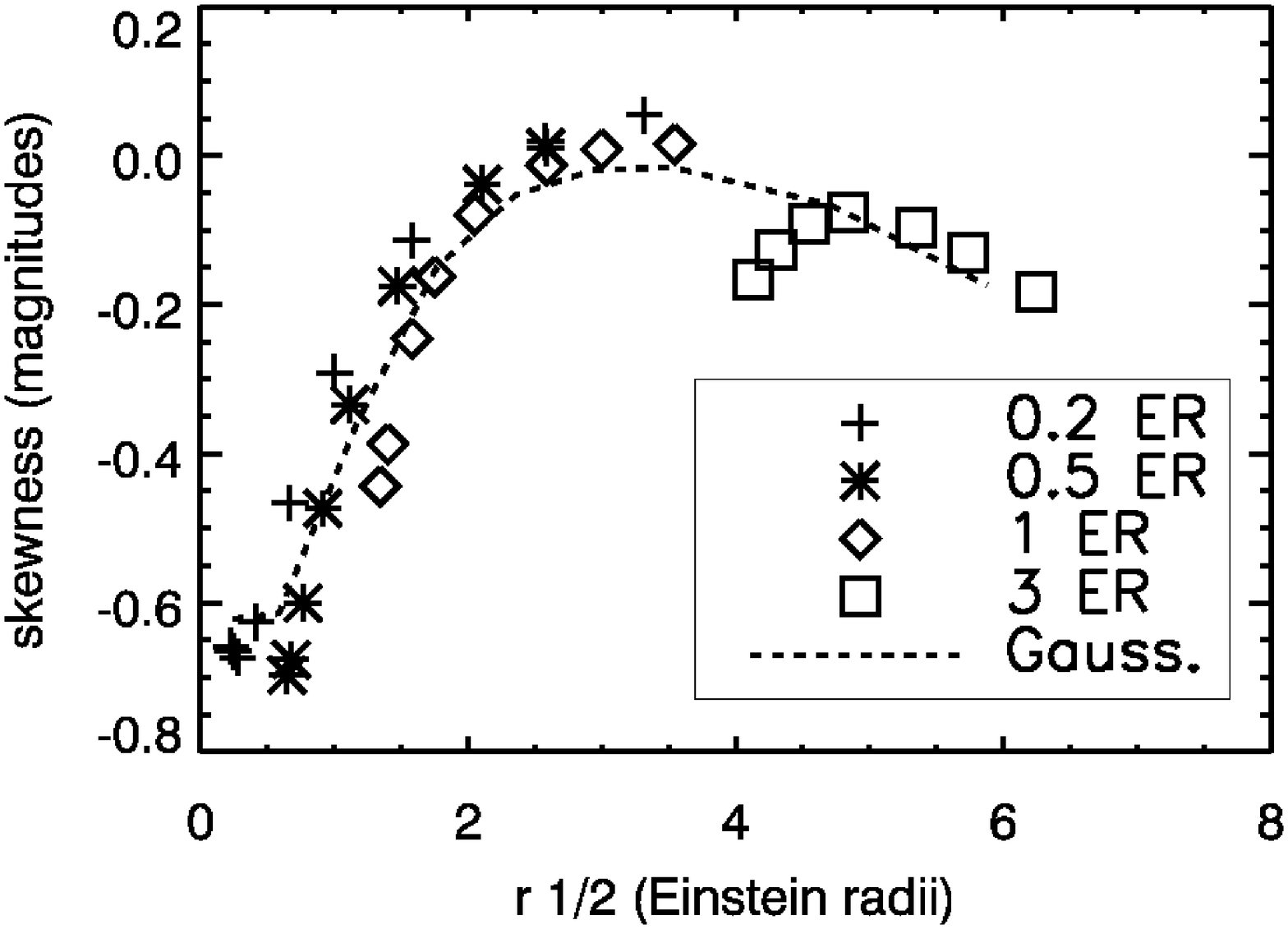}
\caption{Same as Figure~\ref{stat2}, here for the negative parity case $\kappa=\gamma=0.6$.\label{stat3}} 
\end{figure}

By plotting dispersion and skewness against half-light radius 
(Figures~\ref{stat2} and~\ref{stat3}),
we discovered that for any particular disk model there is a clear dependence of
dispersion and skewness on the half-light radius, 
but if we compare disk models with different 
shapes but the same half-light radius, the dispersion and skewness of the associated
histograms are only slightly dependent on the shape of the model.  This suggests that
size differences have a more significant effect on microlensing fluctuations than
shape differences do, at least for circular sources.

The chi-square tests confirm this result. When comparing magnification histograms, the 
number of observations
needed to distinguish sources with differently-shaped brightness profiles but the same
size
is significantly higher than the amount needed to tell the difference between sources
with different sizes but the same shape of the brightness distribution.

This is strong
evidence that the dependence of microlensing variability on source shape is far
weaker than the dependence on source size.  We can model the accretion disk by
any circular brightness profile we like---Gaussian disk, uniform disk, or any other
well-behaved disk model---and our model will produce the correct results, as long as
it is the correct size.

Since the physically-motivated disk model we studied was larger than what one 
would expect to observe, further studies of smaller, more typical 
Shakura-Sunyaev disks could help clarify the validity of these conclusions.  
However, our results are valid in the limits of extremely small microlenses or 
large black holes, and we can conclude in general that if any physical properties 
of a disk have an effect on the microlensing of quasars away from caustics, it 
is the half-light radius of the source and not the shape of its brightness 
profile. 

\acknowledgments

The authors would like to thank the National Science Foundation for supporting 
this work under grant AST02-06010.  We also thank the referee for asking a 
question that led to important revisions, and Scott Hughes and Roger Blandford for
answering questions that helped implement those revisions.

\clearpage

\clearpage

\clearpage

\clearpage

\clearpage

\clearpage

\clearpage

\clearpage

\clearpage

\clearpage

\clearpage

\clearpage

\end{document}